\documentclass[12pt, draftclsnofoot, onecolumn]{IEEEtran}

\usepackage{graphicx,cite,acronym,amsmath,amssymb,amsfonts,color,floatflt,stfloats}
\usepackage{caption}
\usepackage{subcaption}
\usepackage{epsfig,epstopdf,psfrag}
\usepackage[ruled,vlined]{algorithm2e}
\SetKwInput{KwInput}{Input}                
\SetKwInput{KwOutput}{Output}  

\usepackage[table]{xcolor}
\usepackage{bm}

\usepackage[cmintegrals]{newtxmath}
\usepackage{graphics,hhline,array,cite,bbm}

\usepackage{latexsym}
\usepackage{theorem}
\usepackage{times}
\usepackage{makeidx}

\DeclareGraphicsExtensions{.png,.jpg,.eps}

\acrodef{WMMSE}{weighted minimum mean square error}

\acrodef{IRS}{intelligent reflecting surface}
\acrodef{EM}{electromagnetic}
\acrodef{AR}{achievable rate}

\acrodef{MU-MISO}{multi-user MISO}
\acrodef{SU-MIMO}{single-user MIMO}
\acrodef{MU-MIMO}{multi-user MIMO}
\acrodef{MIMO}{multiple-input multiple-output}
\acrodef{MISO}{multiple-input single-output}
\acrodef{SISO}{single-input single-output}
\acrodef{OFDM}{orthogonal frequency division multiplexing}

\acrodef{mmWave}{millimeter wave}

\acrodef{GC}{global constraint}
\acrodef{LC}{local constraint}

\acrodef{CR}{channel response}

\acrodef{BS}{base station}

\acrodef{MS}{mobile station}

\acrodef{UE}{user equipment}

\acrodef{RIS}{reconfigurable intelligent surface}

\acrodef{LIS}{large intelligent surface}

\acrodef{MIS}{medium intelligent surface}

\acrodef{SIS}{small intelligent surface}

\acrodef{CSI}{channel state information}

\acrodef{RV}{random variable}
\acrodef{i.i.d.}{independent, identically distributed}
\acrodef{PSD}{power spectral density}
\acrodef{PDF}{probability distribution function}
\acrodef{pdf}{probability distribution function}
\acrodef{CDF}{cumulative distribution function}

\acrodef{AWGN}{additive white Gaussian noise}

\acrodef{RSSI}{received signal strength indicator}

\acrodef{SNR}{signal-to-noise ratio}
\acrodef{SINR}{signal-to-interference-noise ratio}

\acrodef{LOS}{line-of-sight}
\acrodef{NLOS}{non-line-of-sight}

\acrodef{MAC}{medium access control}
\acrodef{RSS}{received signal strength}
\acrodef{MF}{matched filter}
\acrodef{ML}{maximum likelihood}
\acrodef{MSE}{mean square error}
\acrodef{MMSE}{minimum-mean-square-error}
\acrodef{RMSE}{root mean square error}
\acrodef{LS}{least squares}

\acrodef{MUI}{multi-user interference}
\acrodef{PDP}{power delay profile}

\acrodef{RF}{radiofrequency}

\theoremstyle{plain}
\newtheorem{thm}{Theorem}

\DeclareMathOperator{\tr}{tr}

\DeclareMathOperator{\st}{subject\ to}
\DeclareMathOperator{\E}{\mathbb{E}}

\usepackage{bbm}

\newcommand{\0}{\mathbbm{0}}

\newcommand{\mbf}{\mathbf}

\newcommand{\om}{\text{\boldmath{$\omega$}}}
\newcommand{\op}{\left(\text{\boldmath{$\omega$}}\right)}

\newcommand{\Gt} {G_{\text{T}}}
\newcommand{\Gr} {G_{\text{R}}}

\newcommand{\thetai} {\theta_{\text{inc}}}
\newcommand{\phii} {\phi_{\text{inc}}}

\newcommand{\Gc} {{  G_{\text{c}}}}

\newcommand{\Acc} {{  A_{\text{c}}}}

\newcommand{\Thetai} { \Theta_{\text{inc}}}

\IEEEoverridecommandlockouts

\usepackage{tikz}

\usetikzlibrary{positioning} 
\definecolor{Eored}{rgb}{.647 ,.129 ,.149} 
\definecolor{Eogreen}{rgb}{0 ,0.53 ,0}

\begin{document}

\title{Intelligent Reflecting Surfaces: Sum-Rate Optimization Based on Statistical CSI}

\author{Andrea~Abrardo, \IEEEmembership{Senior~Member,~IEEE}, Davide~Dardari, \IEEEmembership{Senior~Member,~IEEE}, and Marco~Di~Renzo, \IEEEmembership{Fellow,~IEEE}
\vspace{-1.50cm}
\thanks{Manuscript received Dec. 19, 2020. A. Abrardo is the Univerity of Siena and CNIT, Italy (e-mail: abrardo@dii.unisi.it). D. Dardari is with the University of Bologna and CNIT, Italy (e-mail: davide.dardari@unibo.it). M. Di Renzo is with CNRS and Paris-Saclay University, France (e-mail: marco.di-renzo@universite-paris-saclay.fr).}
}

\maketitle

\begin{abstract}
In  this  paper, we  consider  a multi-user multiple-input multiple-output (MIMO) system aided by multiple intelligent reflecting surfaces (IRSs) that are deployed to increase the coverage and, possibly, the rank of the channel. 
We propose an optimization algorithm to configure the IRSs, which is aimed at maximizing the network sum-rate by exploiting only the statistical characterization of  the  environment,  such  as  the distribution of the locations   of  the  users  and the distribution of the multipath channels. As a consequence,  the proposed approach does not require the estimation of the instantaneous  channel state information (CSI) for system optimization, thus significantly  relaxing (or even avoiding)  the  need  of frequently reconfiguring the IRSs, which constitutes one of the most critical issues in IRS-assisted systems.    
Numerical results confirm the validity of the proposed approach. It is shown, in particular, that IRS-assisted wireless systems that are optimized based on statistical CSI still provide large performance gains as compared to the baseline scenarios in which no IRSs are deployed.
\end{abstract}

\section{Introduction}
\label{Sec:Introduction}

The path towards sixth generation (6G) wireless networks foresees the use of new frequency bands such as the terahertz (THz) spectrum, as well as the development of new transmission technologies in order to comply with the challenging requirements introduced by new mobile applications \cite{DavBer:18,San:19,Zha:19,RapXinKanJuMadManAlkTri:19}.
Among several promising transmission technologies, one technology that has recently received a significant attention is the \ac{IRS} \cite{LiaNieTsiPitIoaAky:18}. An \ac{IRS} can be broadly defined as a man-made thin surface, which can be realized by using inexpensive antenna elements or metamaterials, whose properties provide great flexibility on how the \ac{EM} waves that impinge upon the surface are reflected, refracted, focused, etc. \cite{AsaAlbTcvRubRadTre:16, HolKueGorOHaBooSmi:12,RadAsaTre:14,Bur:16}. The most recent research activities on \ac{IRS} are concerned with the design of nearly passive reconfigurable planar structures, i.e., surfaces, whose surface impedance can be appropriately configured according to specified communication needs \cite{HolMohKueDie:05,Fink:14,Tur:14,DiRenzo:2019d}. These solutions are referred to as reconfigurable intelligent surfaces (RISs), and are capable of realizing multiple wave transformations (not only anomalous reflection, refraction, or focusing) \cite{DiRenzo2020_JSAC}.

\ac{IRS}s have several applications, and, in particular, they are useful for enhancing the coverage in severe \ac{NLOS} channel conditions by acting as intelligent mirrors \cite{DiRDanXiDeRTre:2020}. More in general, \ac{IRS}s are viewed as a technology enabler for realizing the so-called  \emph{smart radio environment}, i.e., a wireless system in which the environment (e.g., the channel) becomes a variable that can be optimized in addition to the parameters of the communication devices \cite{DiRenzo:19b}. 
For instance, an \ac{IRS} can be programmed to create ``artificial" multipath in such a way that the channel rank, and so the channel capacity, in \ac{MIMO}-aided systems is increased \cite{OzdBjoLar:20,Dar:J20}. In the presence of multiple users, \ac{IRS}s can be optimized to maximize the \ac{SINR} or the \ac{AR} \cite{HuRusEdf:18}.

One of the main design and implementation challenges of \ac{IRS}-assisted communication systems is the need of estimating the \ac{CSI} of the transmitter-\ac{IRS} and \ac{IRS}-receiver channels, and the need of appropriately configuring (virtually in real-time) the operation of the \ac{IRS} via a feedback channel. As  recently shown in \cite{Zappone2020},  the overhead associated with the channel estimation and reporting phases through a feedback channel may offset the performance gains introduced by an \ac{IRS}, if the number of individually reconfigurable elements is too large and/or an inefficient channel estimation algorithm is employed. Furthermore, due to the dynamic behavior of the wireless channel and the mobility of the users, an \ac{IRS} needs to have the capability of re-configuring itself at a time-scale that depends on the coherence time of the environment (e.g., the channel), which may be difficult to realize in some highly dynamic environments \cite{WuZhaZheYouZha:20}.

Motivated by these considerations, there is general a consensus that it is imperative to develop low-overhead \ac{IRS}-assisted systems, which are designed and optimized without relying on the perfect knowledge of the \ac{CSI}, while still providing performance gains with respect to legacy wireless networks in the absence of \ac{IRS}s. The present paper tackles this open research issue, by proposing a \textit{joint offline and online optimization design} for \ac{IRS}-assisted communication systems. In the next two sub-sections, we first review the research works related to ours and then summarize the main contributions and novelty of the proposed approach.

\subsection{State of the art}
A comprehensive state of the art review on \ac{IRS}-assisted systems can be found in \cite{DiRenzo2020_JSAC}, \cite{WuZhaZheYouZha:20}, \cite{Yuanwei2020}. The present paper is focused on optimizing IRSs under the assumption that the perfect knowledge of the \ac{CSI} is not available. In this section, therefore, we discuss only research works that are closely related to this main focus of the paper. Interested readers are referred to  \cite[Sec. V-J]{DiRenzo2020_JSAC}, \cite[Table III]{WuZhaZheYouZha:20} and  \cite[Sec. V-G]{DiRenzo2020_JSAC}, \cite[Table V]{WuZhaZheYouZha:20} for a detailed summary of the available research results on resource allocation and optimization, and channel estimation, respectively.

Based on the analysis of the state of the art (see, e.g., \cite[Table III]{WuZhaZheYouZha:20}), we evince that the vast majority of research works focused on optimizing the performance of \ac{IRS}-assisted systems are based on the perfect knowledge of the \ac{CSI} for all the available channels, as well as on the availability of appropriate feedback mechanisms to configure the operation of the \ac{IRS}s based on the acquired \ac{CSI}. This may make the channel estimation and signaling (feedback) overhead prohibitively high, especially in \ac{MU-MIMO} systems \cite{Zappone2020}. More precisely, several channel estimation schemes have been proposed for different system configurations (single/multiple users, single/multiple antennas), as recently surveyed in \cite[Sec. V-G]{DiRenzo2020_JSAC}. The common denominator of most of the available schemes is to let the \ac{IRS}s change the analog beamforming vectors during the channel estimation phase according to pre-designed reflection patterns that are exploited and appropriately optimized to facilitate and to make robust the channel estimation process. In spite of the relatively large number of research contributions available to date, the channel estimation in \ac{IRS}-aided systems is still an open issue that is characterized by three major challenges: (i) the long training time, especially in \ac{MU-MIMO} systems, which may not be tolerable in dynamic scenarios; (ii) the real-time reconfiguration of the reflection functionality of the \ac{IRS} through a dedicated control channel with the \ac{BS}; and (iii) the need of ad hoc channel estimation and signaling protocols that make the deployment of \ac{IRS}s non-transparent to existing communication protocols. 
In order to tackle these open research issues, some authors have recently started researching on \ac{IRS}-aided systems that do not necessarily rely upon the perfect knowledge of the \ac{CSI}, e.g., \cite{Nadeem2020}, \cite{WuZhang2020}, \cite{Chaaban2020}, \cite{Miaowen2020}, \cite{Cunhua2020}, \cite{DarMas:20}. 

In \cite{Nadeem2020}, the authors formulate an optimization problem for determining the optimal linear precoder, the power allocation matrix at the transmitter, and the phase shifts matrix at the IRS that maximize the minimum \ac{SINR} of the system. The authors develop a deterministic
approximation for the parameters of the optimal linear precoder by using tools from random matrix theory. Based on the obtained deterministic approximations, an algorithm based on the projected gradient ascent method is proposed to solve the non-convex optimization problem for determining
the phases that maximize the minimum \ac{SINR} of the users. The developed algorithm requires the knowledge of only the large-scale channel statistics, including the slowly varying spatial correlation matrices, which can be estimated using the knowledge of only the mean angles and angular spreads, and the channel attenuation coefficients, which change slowly with time.
A \ac{MISO} downlink multiuser communication system is considered in \cite{WuZhang2020}, in which the phase shifts of a single \ac{IRS} and the transmit beamforming/precoding vector at the access point are optimized by relaying on a two-timescale beamforming optimization algorithm. More precisely, the phase shifts applied by the \ac{IRS} are optimized based on the statistical \ac{CSI} of all the links, and the transmit beamforming/precoding vectors at the access point are designed by relying on the instantaneous \ac{CSI} of the effective channels, given the optimized configuration of the \ac{IRS}. The authors show that the proposed approach can reduce the channel training overhead and the complexity of the analog beamforming design over existing schemes that need the knowledge of the instantaneous \ac{CSI} of all the available channels. It is shown that, even though the phase shifts are optimized by using statistical \ac{CSI}, a significant performance gain in terms of sum-rate can be obtained, as compared with systems in the absence of an \ac{IRS}.
The authors of \cite{Chaaban2020}  study an IRS-assisted \ac{SISO}  broadcast channel, in which the \ac{IRS} elements introduce random or deterministic phase rotations without requiring instantaneous \ac{CSI}. The
average sum-rate capacity of the considered system is achieved by opportunistic scheduling the user with the largest \ac{SNR}. Also, the authors show that the average sum-rate under the proposed scheme in a slow-fading environment is capable of approaching the performance of coherent beamforming as the number of users increases. 
In \cite{Miaowen2020}, the authors propose an \ac{IRS}-enhanced \ac{MISO}  system with reflection pattern modulation, where the \ac{IRS} can configure its reflection state for enhancing the received signal power via analog beamforming and simultaneously conveying its own information via reflection modulation. The authors formulate an optimization problem to maximize the average received signal power by jointly optimizing the digital beamforming at the access point and the analog beamforming at the \ac{IRS} by assuming that the state information at the \ac{IRS} is statistically known by the access point, and that the \ac{CSI} is affected by channel estimation errors. A sub-optimal algorithm for solving the resulting optimization problem based on the alternating optimization method is proposed as well.
The authors of  \cite{Cunhua2020} investigate an \ac{IRS}-aided multi-pair communication system, in which multi-pair users exchange information via an \ac{IRS}. The authors derive an approximated expression of the achievable rate by assuming that only the knowledge of the statistical \ac{CSI} is available.
In  \cite{DarMas:20}, finally, the estimation of the local \ac{CSI} and the need for a dedicated control channel are avoided by exploiting the frequency selectivity of the \ac{IRS}s. It is shown that, if the \ac{IRS}s are appropriately designed, different reflection properties can be obtained in an \ac{OFDM} context by suitably changing the sub-carriers allocated to each user. The effectiveness of the method is validated in the far-field and near-field regimes.

\subsection{Contribution of the paper}

In the present paper, we consider a multi-\ac{IRS} \ac{MU-MIMO} system and we optimize the reconfigurable elements of the \ac{IRS}s and the beamforming vectors at the BS and \acp{UE} so as to maximize the system sum-rate. The main novelty of the proposed approach lies in not requiring the instantaneous \ac{CSI} for optimizing the IRSs, thus relaxing the need for their (real time) configuration. As far as the optimization of the IRSs is concerned, in fact, the proposed approach relies only on the statistical characterization of the environment, such as the distribution of the users' locations and the channel statistics. To this end, a two-phase optimization process is introduced, which encompasses an \textit{offline} phase (long-term and sporadic) and an \textit{online} (short-term and more frequent) phase. During the offline phase, the IRSs are optimized. During the online phase, the BS and the UEs are optimized. The main advantage of the proposed approach is that the statistical \ac{CSI} can be either known a priori (e.g., one may know that some UEs are confined within a certain area) or can be learned occasionally during the operation of the system. {In the former case, the IRSs need to be optimized once forever during the offline optimization phase, and no dedicated control channel is needed during the communication phase.} In the latter case, the IRSs need to be optimized sporadically based on the \textit{distribution} of the users' locations and the fading channels. In this case, the configuration of the \ac{IRS} is not kept fixed forever but it is updated at a low rate and requires a low estimation and feedback overhead. As far as the online phase is concerned, on the other hand, the \ac{IRS} is always kept fixed and is not part of the optimization process. During the online phase, the \ac{IRS} is viewed as a \textit{smart scatterer} that generates controlled multipath based on the statistical distribution of the channels and the users' locations. Hence, the \ac{IRS} is implicitly embedded in the channel, and only the BS and the UEs need to be optimized by using conventional \ac{CSI} acquisition and beamforming methods \cite{Abrardo2019}.

To substantiate the proposed approach, numerical results are illustrated for analyzing the coverage and channel rank improvement of the system sum-rate as a function of the statistical distribution of the users' locations, as well as the number of UEs and \ac{IRS}. Monte Carlo simulations show that, even in the presence of moderate a priori knowledge about the scenario, the proposed approach leads to a significant performance improvement when the \ac{LOS} is not available. Moreover, the numerical results indicate that increasing the number of \ac{IRS}s deployed in the environment, while keeping  constant the total surface area, increases the effective rank of the channel and reduces the interference at the \acp{UE}. These results make the proposed approach an appealing solution to facilitate the deployment of \ac{IRS}s in wireless systems.

\subsection{Comparison with the state of the art}

Compared with the related research works discussed in Section I-B \cite{Nadeem2020}-\!\!\cite{DarMas:20}, the most closely related paper to ours is \cite{WuZhang2020}. Therein, an IRS-assisted \ac{MISO} system is considered, in which the beamforming optimization problem is split into a short-term and a long-term optimization phases. In the first phase, the IRS analog beamforming vector is assumed to be fixed and a conventional \ac{WMMSE} algorithm is utilized to compute the \ac{BS} digital beamforming vector with the goal of maximizing the system sum-rate. In the second phase, the \ac{IRS} analog beamforming vector is optimized with the goal of maximizing the rate averaged with respect to the small-scale fading statistics. These latter statistics, together with the \ac{LOS} channel coefficients, are estimated during a preceding transmission phase in which the \ac{IRS} is in sensing mode and pilots and/or data symbols transmitted in the uplink and downlink are exploited for channel estimation. The optimization is performed by leveraging the close relationship between the rate and the \ac{SNR} in \ac{MISO} systems, and by exploiting such a relationship to derive the gradient of each user rate with respect to the IRS analog beamforming vector. 

In the present paper, on the other hand, we consider a general multi-\ac{IRS} and \ac{MU-MIMO} communication system, in which the digital beamforming vectors at the BSs and UEs and the analog beamforming vectors at the IRSs are optimized with the goal of maximizing the average sum-rate. In addition, rather than considering a long-term optimization problem, we consider an \textit{offline}-based approach in which only mild a priori knowledge on the network scenario is available. More specifically, we assume the a priori knowledge of only the statistical distribution of the users' locations in a given service area. This allows us to get rid of estimating the exact \ac{LOS} channel coefficients. Accordingly, the \ac{IRS}s can be assumed to be nearly-passive and do not need to be equipped with any ad hoc circuits (or sensors) for estimating the \ac{CSI}. 

As far as the calculation of the average sum-rate is concerned, similar to \cite{WuZhang2020}, we resort to a Monte Carlo generation method for the channels. However, we include the generation of the users' locations as well. By capitalizing on the proposed approach, in the extreme case in which the network is static as a function of the users' locations and channel statistics, the IRSs need to be optimized only once during the offline phase. In this case, the IRSs act as passive repeaters.

As far as the optimization algorithm is concerned, we provide new contributions for the online and offline optimization phases. As for the online phase, we introduce a generalized version of the \ac{WMMSE} algorithm for application to \ac{MU-MIMO} systems. As for the offline phase, owing to the more involved expression of the per-user rate of \ac{MIMO} systems with respect to \ac{MISO} systems, we further generalize the \ac{WMMSE} method for optimizing the analog beamforming vectors of the \ac{IRS}s. It is worth noting that the \ac{WMMSE} algorithm was recently used for optimizing \ac{IRS}s \cite{Pan2020}. Therein, however, the optimization is performed for a given instance of the channel (the instantaneous \ac{CSI} is needed) and a single \ac{IRS} is considered. 

Finally, we emphasize that the terms offline and online optimization were recently used in \cite{NajMarJamSchPoo:20}. Therein, however, the problem formulation and proposed approach are fundamentally different compared with ours. In \cite{NajMarJamSchPoo:20}, for example, during the online optimization phase, all the effective end-to-end channels between the transmitter and the user via each tile of the \ac{IRS}s are needed. This does not apply to the proposed algorithm, since the individual knowledge of these channels is not necessary. Similar to \cite{NajMarJamSchPoo:20}, however, our proposed approach can be applied to \acp{IRS} that are made of multiple tiles, which is useful for reducing the optimization complexity.

\subsection{Organization of the paper}

This paper is structured as follows. In Section, \ref{sec:ChannelModel} the channel and \ac{IRS} physical models are introduced. In Section \ref{Sec:opt}, the offline and online \ac{IRS} optimization problems are described and tackled. Numerical results referred to a typical indoor scenario are described in Section \ref{sec:NumericalResults}.
Finally, concluding remarks are given in Section \ref{sec:Conclusion}.

\section{Direct and \ac{IRS}-Assisted Channel Models}
\label{sec:ChannelModel}

We consider the downlink of a wireless system in which one \ac{BS} equipped with $M$ antennas serves $N_u$ \acp{UE} that comprise $L$ antennas each. In this network scenario, $N_{IRS}$ \acp{IRS} are deployed in some predefined locations for assisting the communication between the \ac{BS} and the \acp{UE}. Each \ac{IRS} comprises $P_{IRS}$ nearly-passive reconfigurable scattering elements. In this setting, each IRS can be conveniently partitioned into $K_t$ smaller tiles in order to reduce the design complexity (e.g., see \cite{NajMarJamSchPoo:20}). In this case, the number of equivalent IRSs is $K = N_{IRS} \times K_t$, each comprising $P = P_{IRS}/K_t$ scattering elements. In the following, for ease of notation, we employ the notation $K$ and $P$ regardless of whether the IRSs are partitioned in tiles or not.

The objective of this section is to introduce the \ac{BS}-\ac{UE} and \ac{BS}-\ac{IRS}-\ac{UE} channel models.  For ease of writing, we formulate the channel models for a single \ac{UE}.

\subsection{Channel model for the direct link (\ac{BS}-\ac{UE} channel)}\label{direct_link}
 
We consider a non-stationary narrow-band clustered channel model, typically used for application to \ac{mmWave} communications \cite{Shangbin2015}, which accounts for the near-field and far-field effects of the \ac{EM} propagation in the presence of large antenna arrays. We denote by $f_0$ the center frequency of the \ac{EM} signal and by $\lambda$ the corresponding wavelength. The channel is composed of a \ac{LOS} component and a \ac{NLOS} component that comprises a number of clustered paths, each corresponding to a macro-level scattering path. In particular, we assume that the center of each cluster $\mathbf{c}_q$ of multipath is located on an ellipse whose foci coincide with the antennas of the \ac{BS} and \ac{UE}. For simplicity, we assume that $\mathbf{c}_q$ are all located on the same ellipse. By denoting with $\mathbf{t}$ and $\mathbf{r}$ the center of the antenna arrays of the \ac{BS} and \ac{UE}, respectively, we have, by definition of ellipse, $\left|\mathbf{c}_q-\mathbf{t}\right|+\left|\mathbf{c}_q-\mathbf{r}\right| = \left|\mathbf{t}-\mathbf{r}\right|/\epsilon$, $\forall q$, $\epsilon$ being the ellipse eccentricity. Each cluster consists of a number of paths, and each path is represented by a point, which is denoted by $\mathbf{p}_{ql}$, that is randomly placed within a maximum (small) distance from the center of the cluster $\mathbf{c}_q$. Accordingly, all the paths of a given cluster reach the receiver within a limited angular spread, in agreement with typical \ac{mmWave} channels.  
In mathematical terms, the channel complex gain can be formulated as
\begin{equation} \label{eq:chmodel1}
h_{mn}=X_d \, \beta_{mn}^{\text{(LOS)}}\, 
\exp \left (-\jmath \frac{2 \pi \left |\mathbf{t}_m-\mathbf{r}_{n}\right|} {\lambda} \right ) 
+ \frac{\beta^{\text{(NLOS)}}_{nm}}{N_pN_c} \, \sum\limits_{q=1}^{N_{c}}\sum\limits_{l=1}^{N_{p}} \alpha_{ql}  \, \, \exp \left (-\jmath 2\pi\frac{\left|\mathbf{t}_m-\mathbf{p}_{ql}\right|+\left|\mathbf{r}_n-\mathbf{p}_{ql}\right|}{\lambda} \right )
\end{equation}
where $\mathbf{t}_m$ denotes the position of the $m$th antenna-element of the \ac{BS}, $\mathbf{r}_n$ denotes the position of the $n$th antenna-element of the \ac{UE},  $\beta_{mn}^{\text{(LOS)}}=\frac{\sqrt{\Gt \Gr} \lambda}{4\pi \left |\mathbf{t}_m-\mathbf{r}_{n}\right|}$ denotes the path-loss gain of the \ac{LOS}, $\Gt$ denotes the antenna-element gain at the \ac{BS}, $\Gr$ denotes the antenna-element gain at the \ac{UE}, $N_c$ denotes the number of clusters, $N_p$ denotes the number of paths of each cluster, $\alpha_{ql} \sim \mathcal{CN}\left(0,1\right)$ denotes the random complex fading coefficient  of the $l$th path of the $q$th cluster, and $\beta^{\text{(NLOS)}}_{nm}$ denotes the path-loss gain of the channel in \ac{NLOS}. 
According to \cite{WhitePaperChannel}, we have $\beta^{\text{(NLOS)}}_{nm}\, \text{(dB)}=-\text{PL}_0 - 10\, n_e \, \log_{10} d$, where $\text{PL}_0$ denotes the path-loss (in dB) at the reference distance of one meter, $d$ denotes the distance between the center of the antenna-array of the \ac{BS} ($\mathbf{t}$) and the center of the antenna-array of the \ac{UE} ($\mathbf{r}$), and $n_e$ denotes the path-loss exponent. In particular, we consider the \emph{ABG} channel model for the indoor office (IO) and the shopping mall (SM) \ac{NLOS} scenairos at 28 GHz \cite{WhitePaperChannel}. The main difference between the two scenarios is given by the path-loss exponents, which are equal to 3.83 and 3.21 for the IO and SM cases, respectively. Finally, the binary variable $X_d$ allows us to model whether the \ac{LOS} is present ($X_d=1$) or is obstructed ($X_d=0$). 
 

\subsection{Reflection model for the \ac{IRS}}
\label{sec:IRSmodel}
We consider an \ac{IRS} that comprises $P$ small unit cells (scattering elements), which are spaced $\lambda/2$ apart and whose area is $\Acc$. We consider a locally-plane (with respect to the size of each unit cell)\footnote{The wave is assumed to be plane with respect to the area $\Acc$, but it is spherical with respect to the size of the \ac{IRS}.} incident wave,
whose angle of incidence with respect to the IRS is $\Thetai=(\thetai,\phii)$, with $\phii \in [0,2\pi)$ and $\thetai \in [0,\pi)$ denoting the azimuth and elevation angles, respectively.

The local reflection coefficient of the $p$th unit cell towards the generic direction of scattering $\Theta=(\theta,\phi)$ is modeled as 
\begin{align} \label{eq:r}
 r_p(\Thetai,\Theta)=\sqrt{F(\Thetai) \, F(\Theta)} \, \Gc \, b_p 
\end{align}
where $F(\Theta)$ denotes the normalized power radiation pattern of each unit cell, which accounts for the possible non-isotropic behavior of each scattering element. As a first approximation, we assume that $F(\Theta)$ is frequency-independent within the bandwidth of interest. Also, $\Gc$ denotes the boresight gain of the unit cell, and $b_p=\rho_p \exp (\jmath \phi_p)$ denotes the load reflection coefficient, with $\rho_p=|b_p|$ and $\phi_p$ being the amplitude and the phase shift introduced by the $p$th unit cell, respectively.\footnote{Depending on the specific implementation, the load reflection coefficient may have a different physical meaning \cite{Gradoni}.} For analytical convenience, we collect the load reflection coefficients $b_p=\rho_p \exp (\jmath \phi_p)$ in the vector $\mathbf{b}=[b_1,b_2, \ldots, b_P]$.
It is worth mentioning that, especially for large \acp{IRS}, the angle of incidence $\Thetai$ can be different for each unit cell. 
Without loss of generality, the boresight gain is defined as
$\Gc=\Acc\,  4\pi /\lambda^2$. In addition, a quite general model for $F(\Theta)$ is 
\begin{align} \label{eq:F}
    F(\Theta)=\left\{\begin{array}{ll} \cos^q (\theta)  & \, \, \, \, \, \theta \in [0,\pi/2] \, , \phi \in [0,2\pi]  \\0 & \, \, \, \, \, \text{otherwise}  \\ \end{array}\right.   
 \end{align}
which is compliant with the models used in, e.g., \cite{Ell:19,TanCheDaiHanDiRZenJinCheCui:19,OzdBjoLar:20} for different values of the parameter $q$, and with the model recently derived in \cite{NajMarJamSchPoo:20}, which originates from the radar cross-section of small reflecting tiles. In the numerical results, without loss of generality, the following parameters are considered: $q=0.57$, $\Acc=\frac{\lambda^2}{4}$ \cite{Ell:19,DarMas:20}.

\subsection{Channel model for the indirect link (BS-IRS-UE channel)}

In \ac{IRS}-assisted communications, the near-field effects cannot be ignored if the size of the \ac{IRS} is large and the transmission distances are short \cite{MDR_ChannelRIS}. In this case, the plane wave approximation is not fulfilled anymore, and spherical waves need to be considered. Thus, the channel is non-stationary along the \ac{IRS}. This implies that the angles of incidence and scattering are not sufficient to characterize the global response of the \ac{IRS}, and the different transmission distances between the unit cells of the \ac{IRS} and the antennas of the \ac{BS} and \ac{UE}s need to be taken into account.

In general, the \ac{IRS}s are strategically deployed in order to ensure that they are in \ac{LOS} with the \ac{BS} and the \ac{UE}s. In addition, they are more useful when the \ac{BS} and the \ac{UE}s are in \ac{NLOS} \cite{Perovic}. Therefore, these assumptions are considered in the present paper and in the numerical results. In addition, second-order reflections between the \ac{IRS} and the cluster of scatterers are ignored because they are expected not to contribute significantly due to the large path-loss experienced.

According to the \ac{IRS} model in \eqref{eq:r}, the (complex) channel gain between the $m$th transmit antenna of the \ac{BS} and the $p$th unit cell of the \ac{IRS} can be written as 
\begin{equation} \label{eq:skj}
s_{pm} = \frac{\sqrt{\Gt \Gc \, F(\Theta_{pm})} \lambda}{4\pi d_{pm}} 
\exp \left (-\jmath \frac{2 \pi d_{pm}} {\lambda} \right )  
\end{equation}
where $d_{pm}$ and $\Theta_{pm}$ denote the distance between the $m$th transmit antenna of the \ac{BS} and the $p$th unit cell of the \ac{IRS}, and the angle from which the $m$th transmit antenna is viewed by the \ac{IRS}, respectively.
Similarly, the  (complex) channel gain between the $p$th unit cell of the  \ac{IRS}  and the $n$th receive antenna can be written as
\begin{align} \label{eq:tik}
t_{np} &= \frac{\sqrt{\Gr \Gc \, F(\Theta_{np})} \lambda}{4\pi d_{np}} \exp \left (-\jmath \frac{2 \pi d_{np}} {\lambda} \right )   
\end{align}
where $d_{np}$ and $\Theta_{np}$ denote the distance between the $n$th receive antenna and the $p$th unit cell of the \ac{IRS}, and the angle from which the $n$th receive antenna is viewed by the \ac{IRS}, respectively.

It is important to note that, differently from \eqref{eq:chmodel1}, the \ac{NLOS} component of the channel is not present in \eqref{eq:skj} and \eqref{eq:tik}, since it is related to the second- and higher-order reflections, which are typically negligible, as mentioned in previous text, and are hence ignored in this paper.   

For generality, we consider IRSs that are designed obeying a \ac{LC} and a \ac{GC} design criteria \cite{AsaAlb:16}. In the first case, we assume $\rho_p \le 1\, \, \forall p$, i.e., each unit cell of the IRS scatters an amount of power that is smaller than or equal to the impinging power. In the second case, we assume $\sum_p \rho_p^2 \le P$, i.e., the IRS as a whole scatters an amount of power that is smaller than or equal to the total impinging power. Also, we assume that the IRSs are non-lossy and that the phase shifts applied by the unit cells belong to a finite set, i.e., $b_p \in \mathcal{S}_{N_b} = \left\{1,e^{\frac{j2\pi}{M_l }},e^{\frac{j4\pi}{M_l }},\ldots,e^{\frac{j2(M_l-1)\pi}{M_l }}\right\}$, where $M_l = 2^{N_b}$ is the number of quantization levels.

\section{IRS Optimization Based on Statistical CSI} \label{Sec:opt}

In this section, we formulate the problem statement and introduce the proposed (offline and online) optimization algorithms for sum-rate maximization.

\subsection{Problem formulation}
\label{Sec:model}

Let $\mathbf{s}_j= \left[s_{j}(1),s_{j}(2),\ldots,s_{j}(L)\right]^T$ be the complex vector comprising the $L$ symbols\footnote{We assume that the number of symbols is equal to the number of receive antennas.} transmitted by the BS to the $j$th UE (also referred to as the $j$th stream) and $\mathbf{V}_j \in \mathbb{C}^{M \times L}$ be the corresponding precoding matrix.
The transmitted vector can be formulated as
\begin{equation}
\mathbf{x}_j = \mathbf{V}_j \mathbf{s}_j \quad \in \mathbb{C}^{M \times 1} \, .
\end{equation}

The information symbols are assumed to be zero-mean and \ac{i.i.d.} \acp{RV}, i.e., $\mathbb{E}\left[\mathbf{s}_j \mathbf{s}_j^H\right] =  \mathbf{I}_{L}$ and $\mathbb{E}\left[\mathbf{s}_j \mathbf{s}_i^H\right] =  \mathbf{0}_{L}$ for $j \ne i$.

The signal received at the $k$th IRS can be formulated as 
\begin{equation}
\begin{array}{c}
\mathbf{r}_{k,j} = \mathbf{S}_{k}\mathbf{x}_j
\quad \in \mathbb{C}^{P \times 1}
\end{array}
\label{eq:rx_irs}
\end{equation}
where  $\mathbf{S}_{k} \in\mathbb{C}^{P \times M}$ is the channel matrix between the BS and the $k$th \ac{IRS}, whose elements are modeled according to \eqref{eq:skj}. The signal reflected by the \ac{IRS} is
\begin{equation}
\begin{array}{c}
\mathbf{r'}_{k,j} = \mathbf{B}_{k}\mathbf{r}_{k,j}
\quad \in \mathbb{C}^{P \times 1}
\end{array}
\label{eq:rx_irs2}
\end{equation}
where $\mathbf{B}_{k} = \text{diag}\left(\mathbf{b}_{k}\right) \in \mathbb{C}^{P \times P}$ is the diagonal matrix whose elements are the load reflection coefficients $\mathbf{b}_{k} \in \mathbb{C}^{P \times 1}$ introduced in Section \ref{sec:IRSmodel}.

The signal received at the $i$th \ac{UE} can be expressed as
\begin{equation}
\begin{array}{c}
\mathbf{z}_{i,k,j} = \mathbf{T}_{i,k}\mathbf{r'}_{k,j} = \mathbf{T}_{i,k} \mathbf{B}_{k} \mathbf{S}_{k}\mathbf{x}_j = \mathbf{\tilde{H}}_{i,k} \left(\mbf{b}_{k}\right) \mathbf{x}_j
\quad \in \mathbb{C}^{L \times 1}
\end{array}
\label{eq:rx_irs3}
\end{equation}
where  $\mathbf{T}_{i,k} \in\mathbb{C}^{L \times P}$ is the channel matrix between the $k$th \ac{IRS} and the $i$th receiver, whose elements are modeled according to \eqref{eq:tik}, and $\mathbf{\tilde{H}}_{i,k} \left(\mbf{b}_{k}\right)= \mathbf{T}_{i,k} \mathbf{B}_{k} \mathbf{S}_{k}$.

In addition, the direct signal emitted by the BS and received by the $i$th \ac{UE} is
\begin{equation}
\begin{array}{c}
\mathbf{w}_{i,j} = \mathbf{{\bar{H}}}_{i} \, \mathbf{x}_j
\quad \in \mathbb{C}^{L \times 1}
\end{array}
\label{eq:rx_no_irs}
\end{equation}
where $\mathbf{{\bar{H}}}_{i}$ is the channel matrix between the \ac{BS} and the $i$th UE.

Thus, considering only the $j$th stream and the $k$th IRS, the received signal at the $i$th UE is
\begin{equation}
\begin{array}{c}
\mathbf{y}_{i,k,j} = \mathbf{z}_{i,k,j} + \mathbf{w}_{i,j} + \mathbf{n}_i =  \left(\mathbf{{\bar{H}}}_{i} + \mathbf{\tilde{H}}_{i,k} \left(\mbf{b}_{k}\right) \right) \mathbf{x}_j  + \mathbf{n}_i
\quad \in \mathbb{C}^{L \times 1}
\end{array}
\label{eq:rx_total}
\end{equation}
where $\mathbf{n}_i \in \mathbb{C}^{L \times 1}$ denotes the additive white Gaussian noise with distribution $\mathcal{CN}\left(0,\sigma_i^2 \mathbf{I}_L\right)$. 

Let $\mathbf{{H}}_{i} \left(\mathcal{B}\right) = \mathbf{{\bar{H}}}_{i} + \sum\nolimits_{k=1}^{K}\mathbf{\tilde{H}}_{i,k} \left(\mbf{b}_{k}\right)$ denote the total channel between the BS and the $i$th UE, where $\mathcal{B}=\left\{\mbf{b}_{1},\mbf{b}_{2},\dots,\mbf{b}_{K}\right\}$ is  the set of vectors containing the reflection coefficients  of the $K$ \acp{IRS} (analog beamforming vectors). The signal received at the $i$th \ac{UE} in the presence of $N_u$ concurrent transmitted streams and $K$ \acp{IRS} is
\begin{equation}
\begin{array}{c}
\mathbf{y}_{i} = \sum\limits_{j=1}^{N_u}\left(\sum\limits_{k=1}^{K}\mathbf{z}_{i,k,j} + \mathbf{w}_{i,j} \right) + \mathbf{n}_i  = \mathbf{{H}}_{i} \left(\mathcal{B}\right) \mathbf{x}_i + \mathbf{{H}}_{i} \left(\mathcal{B}\right) \left(\sum\limits_{j=1 \atop j \neq i}^{N_u }\mathbf{x}_j\right) + \mathbf{n}_i
\quad \in \mathbb{C}^{L \times 1} \, .
\end{array}
\label{eq:rx_total}
\end{equation}

Based on the \ac{MIMO} interference channel in \eqref{eq:rx_total}, the \ac{AR} of the $i$th UE is \cite{Shi2011} 
\begin{equation}
\label{eq:rate2.1}
R_{i}(\mathcal{V},\mathcal{B})= \log \det
\left(
{{\bf{I}}_{L}} +
{{\bf{V}}_i^H{\bf{H}}_{i}^H\left(\mathcal{B}\right)\bar{\mbf{J}}_{i}^{ - 1}{\bf{H}}_{i}\left(\mathcal{B}\right)}{{\bf{V}}_i}
\right)
\end{equation}
where $\bar{\mbf{J}}_{i}=\sum\nolimits_{j =1,j \neq i}^{N_u} {{{\bf{H}}_{i}\left(\mathcal{B}\right)}{{\bf{V}}_j}{\bf{V}}_j^H{\bf{H}}_{i}^H\left(\mathcal{B}\right)}  + {\sigma ^2}{{\bf{I}}_{L}}$ is the interference-plus-noise covariance matrix  and $\mathcal{V}=\left\{\mbf{V}_{1},\mbf{V}_{2},\dots,\mbf{V}_{N_{u}}\right\}$ is the set of precoding matrices of all the $N_{u}$ UEs.

In the next two sections, we introduce the proposed approach for optimizing the digital beamforming vectors of the BS and the UEs (which occurs during the online phase), as well as the analog beamforming vectors of the IRSs (which occurs during the offline phase). The online phase is subsequent to the offline phase, but it is introduced first for ease of description. 

\subsection{BS and UEs beamforming optimization -- Online phase}

During the online phase, the analog beamforming vectors $\mathcal{B}$ of the \acp{IRS} are kept constant and are set equal to those computed during the offline phase described next. Hence, for notational simplicity, the dependence on $\mathcal{B}$ of the rate $R_i$ in \eqref{eq:rate2.1} and the channel matrices is omitted in this section. This implies that the impact of the IRSs is incorporated into the channels in a transparent manner and the IRSs operate as smart scatterers. Therefore, the system boils down to a classical MIMO broadcast channel, which can be optimized by relaying on well established methods and protocols for estimating the CSI of the BS-UE links. This case study has been investigated extensively and is, therefore, briefly discussed in this section. In particular, we propose to utilize the iterative WMMSE algorithm for optimizing the digital beamforming vectors of the BS and the UEs \cite{Shi2011}. The WMMSE algorithm is chosen since it provides the basis for optimizing the analog beamforming vectors of the IRSs, which is described in the next section. In preparation for the next section, the main features of the WMMSE algorithm are briefly summarized.

In the online phase, the sum-rate maximization problem can be formulated as
\begin{align}
\label{P:sumRateMax_pre}
&\max \limits_{\mathcal{V}} \sum\limits_{i = 1}^{N_{u}}\alpha_{i}R_{i}\left(\mathcal{V}\right) \\
& \quad \text{subject to} \quad \tr \left(\mbf{V}_i\mbf{V}^H_i \right) \le P_{i}, ~~ i=1,2,\dots,N_{u}  \tag{\ref{P:sumRateMax_pre}.a}\label{Pow_cons}
 \end{align}
where $P_i$ is the power budget of the $i$th UE and  $\boldsymbol{\alpha}=[\alpha_{1},\alpha_{2},\dots,\alpha_{N_{u}}]$ is a set of weights chosen
to guarantee a given degree of fairness among the UEs, e.g., by assigning higher weights to UEs with weaker channels. The readers are referred to \cite{Wang2017} for further information on  $\boldsymbol{\alpha}$. 

The WMMSE algorithm exploits the close relationship between the \ac{SINR} and the \ac{MSE}, in order to find a locally optimal solution to the sum-rate problem in \eqref{P:sumRateMax_pre} \cite{Shi2011}. To elaborate, let us introduce
the MSE matrix $\mbf{E}_{i}\left(\mathcal{V},\mbf{G}_{i}\right)$, for $i=1,2,\ldots,N_u$, as
\begin{equation}
\label{eq:mMSEMat0_1}
\begin{aligned}
\mbf{E}_{i}\left(\mathcal{V},\mbf{G}_{i}\right) & = \E\left\{ {\left( {{{\bf{s}}_i} - {{{\bf{\tilde s}}}_i}} \right){{\left( {{{\bf{s}}_i} - {{{\bf{\tilde s}}}_i}} \right)}^H}} \right\}  \quad \in \mathbb{C}^{L \times L}
\end{aligned}
\end{equation}
where $\mathbf{\tilde s}_i =  \mbf{G}^H_{i} \mathbf{y}_{i}$ and $\mbf{G}_{i}$ is the linear decoding matrix of the $i$th UE.

Problem \eqref{P:sumRateMax_pre} can be reformulated as the following weighted \ac{MSE} minimization problem
\begin{equation}
\label{wMMSE1_1}
\begin{aligned}
&\min \limits_{\mathcal{V},\mathcal{W},\mathcal{G}} \sum\limits_{i = 1}^{N_{u}}\alpha_{i} \big\{\tr\left[\mbf{W}_{i}\mbf{E}_{i}\left(\mathcal{V},\mbf{G}_{i}\right)\right]-\log\det\left(\mbf{W}_{i}\right)\big\}\\
&\quad \st
\quad \tr \left(\mbf{V}_i\mbf{V}^H_i\right)\le P_{i}, ~~ i=1,2,\dots,N_{u}
\end{aligned}
\end{equation}
where $\mbf{W}_{i}\succeq 0$ is the matrix of weights for the MSE of $i$th UE, and $\mathcal{W}=\left\{{\mbf{W}}_{1},\mbf{W}_{2},\dots,\mbf{W}_{N_{u}}\right\}$
and $\mathcal{G}=\left\{\mbf{G}_{1},\mbf{G}_{2},\dots,\mbf{G}_{N_{u}}\right\}$ are the sets of all weight and receive filter matrices, respectively.

The equivalence between problem \eqref{P:sumRateMax_pre} and \eqref{wMMSE1_1} follows by recalling the relation between the MMSE covariance $\hat{\mbf{E}}_{i}\left(\mathcal{V}\right) = \min\limits_{\mbf{G}_{i}}\mbf{E}_{i}\left(\mathcal{V},\mbf{G}_{i}\right)$ and the rate $R_i$, i.e., $R_i = \log \det\left(\hat{\mbf{E}}_{i} ^{-1}\left(\mathcal{V}\right)\right)$ and by the fact that the optimal solution of \eqref{wMMSE1_1} is $\mbf{W}_{i} = \hat{\mbf{E}}^{-1}_{i}\left(\mathcal{V}\right)$. Further details can be found in \cite{Shi2011}.

Problem \eqref{wMMSE1_1} is non-convex. If all the optimization variables are fixed except one, however, it is a convex optimization problem in the remaining variables. Accordingly, the weighted MSE minimization problem in \eqref{wMMSE1_1} can be solved by using an iterative block coordinate descent  (BCD) algorithm \cite{Bertsekas1999}. To elaborate, let us denote by 
$\mbf{G}_{i}^{(q+1)}$, $\mbf{W}_{i}^{(q+1)}$ and $\mbf{V}_{i}^{(q+1)}$ the optimization variables after the $(q+1)$th iteration. Then, the receive filter matrix is \cite{Shi2011}
\begin{equation}\label{MMSEric}
\mbf{G}_{i}^{(q+1)}=  \left(\mbf{J}_{i}^{(q)}\right)^{-1}\mbf{H}_{i}\mbf{V}_{i}^{(q)}
\end{equation}
where $\mbf{J}_{i}^{(q)}=\sum\limits_{j = 1}^{N_{u}} \mbf{H}_{i}\mbf{V}_{j}^{(q)}\left(\mbf{V}_{j}^{(q)}\right)^H\mbf{H}_{i}^H  + \sigma_i^2 \mbf{I}_{L}$, the weight matrix $\mbf{W}_{i}^{(q+1)}$ is
 \begin{equation}
 \label{eq: weight computation}
\mbf{W}_{i}^{(q+1)} =  \left(\mbf{E}_{i}^{(q+1)}\right)^{-1}
\end{equation}
and the precoding filter at the BS $\mbf{V}_{i,q+1}$ is
\begin{equation}
\label{eq:precoderShi_o}
\mbf{V}_{i}^{(q+1)} =\alpha_{i}\left(\mbf{K}^{(q+1)}+ \mu_i \mbf{I}_{M} \right)^{-1}\mbf{H}_{i}^H\mbf{G}_{i}^{(q+1)}\mbf{W}_{i}^{(q+1)}
\end{equation}
where $\mbf{K}^{(q+1)}= \sum\limits_{j = 1}^{N_{u}}\alpha_{j} \mbf{H}_{j}^H \mbf{G}_{j}^{(q+1)} \mbf{W}_{j}^{(q+1)} \mbf{G}_{j}^{(q+1)} \mbf{H}_{j}$ and the Lagrange multiplier $ \mu_i$ involved in the optimization problem is chosen so that the power constraint in \eqref{wMMSE1_1} for the $i$th UE is fulfilled.

\subsection{IRS beamforming optimization -- Offline phase}

The offline phase encompasses the optimization of the \acp{IRS}. In contrast to the online phase that relies on instantaneous CSI, the offline phase relies only on channel long-term statistics, e.g., the distribution of the locations of the UEs in the given area. Therefore, it is executed only sporadically. In the extreme case that the IRSs operate as repeaters, it is executed only once.

The objective of the offline phase is to optimize the analog beamforming vectors of the \ac{IRS}s, $\mathcal{B}$, that maximize the (statistical) average weighted sum-rate of the system. In particular, the average is computed with respect to all random phenomena that determine the channel matrices.

To elaborate, let us denote by ${\Omega}$ the set of random vectors $\text{\boldmath{$\omega$}}$ that determine the channel status, including the locations of the \acp{UE}, the position of the antennas, the distribution of the clusters of multipath for each link, and the complex channel gains. Accordingly, the total channel matrices depend on $\om$ and can be formulated as $\mathbf{{H}}_{i} \left(\om,\mathcal{B}\right)$. Denoting by $f_{\Omega}\left(\text{\boldmath{$\omega$}}\right)$ the joint probability density function of ${\Omega}$, the optimization problem for computing $\mathcal{B}$ can be formulated as
{
\begin{align}
\label{P:sumRateMax1}
&\max \limits_{\mathcal{B}} \mathbb{E}_\om \left [ \max \limits_{\mathcal{V}\op} R_{tot} \left(\mathcal{B},\mathcal{V}\op\right) \right ] = \max \limits_{\mathcal{B}} \int\limits_{\text{\boldmath{$\omega$}} \in \Omega} \max \limits_{\mathcal{V}\op}  \sum\limits_{i = 1}^{N_{u}}\alpha_{i}R_{i}\left(\mathcal{V}\op,\mathcal{B}\right) \, f_{\Omega}\left(\text{\boldmath{$\omega$}}\right) \, d\text{\boldmath{$\omega$}} & \\
& \quad \text{subject to} \notag \\
& \quad \quad \tr \left(\mbf{V}_i\op\ \mbf{V}^H_i\op \right) \le P_{i}, \quad \forall \om, ~~ i=1,2,\dots,N_{u}  \tag{\ref{P:sumRateMax1}.a}\label{Pow_cons}\\
& \quad \quad \tr \left(\mbf{b}_k\mbf{b}_k^H \right) \le 1, \quad k=1,2,\dots,K \tag{\ref{P:sumRateMax1}.b.1}\label{IRS_cons} \quad \quad \quad \quad \quad \quad \text{     (GC case)}\\
& \quad  \text{or} & \nonumber\\
& \quad \quad b_{kp} \in \mathcal{S}_{N_b}, \quad k=1,2,\dots,K \text{  } \quad p=1,\dots,P \tag{\ref{P:sumRateMax1}.b.2}\label{IRS_cons2} \quad  \text{     (LC$_{N_b}$ case)}
 \end{align}
where the dependence of $\mathcal{V}$ with respect to $\om$ is made explicit in order to emphasize that $\mbf{V}_{i}$ depends on $\om$. By direct inspection, we observe that GC in \eqref{IRS_cons} is a convex constraint and \ac{LC}$_{N_b}$ in \eqref{IRS_cons2} is a non-convex constraint. Therefore, the LC setup is more difficult to study. In this section, therefore, we restrict the analysis to the GC setup. After solving the GC case study, a (suboptimal but low complexity) solution for the LC case study can be obtained by recalling that a local element-wise projection onto the unit circle of the solution obtained for the \ac{GC} setup corresponds to the minimum distance projection on the feasibility set of the \ac{LC} setup \cite{Abrardo2019}, \cite{Ghauch2016}. Therefore, we solve the optimization problem in \eqref{P:sumRateMax1} under the \ac{GC} constraint and we then obtain the solution under the \ac{LC} constraint by applying an element-wise scaling to the solution of the GC setup. The performance offered by this approach is analyzed in the numerical results.

Even though the GC constraint is convex, problem \eqref{P:sumRateMax1} is non-convex and is, therefore, difficult to solve. To this end, the sum-rate maximization problem in \eqref{P:sumRateMax1} can be reformulated as a weighted \ac{MSE} minimization problem. In particular, we have
%
\begin{align}
\label{wMMSE1}
\min \limits_{\mathcal{B}}  \int\limits_{\text{\boldmath{$\omega$}} \in \Omega} & \min \limits_{\mathcal{V}\op,\mathcal{W}\op,\mathcal{G}\op} \sum\limits_{i = 1}^{N_{u}}\alpha_{i}  \big\{\tr\left[\mbf{W}_{i}\op\mbf{E}_{i}\left(\mathcal{V}\op,\mathcal{B},\mbf{G}_{i}\op\right)\right]  \nonumber \\
& -\log\det\left(\mbf{W}_{i}\op\right)\big\} f_{\Omega}\left(\text{\boldmath{$\omega$}}\right)  \, d\text{\boldmath{$\omega$}}  \\
\st &\nonumber \\
& \tr \left(\mbf{V}_i \op \mbf{V}^H_i\op \right) \le  P_{i}, \quad  \forall \om , ~~ i=1,2,\dots,N_{u}\\
& \tr \left(\mbf{b}_k\mbf{b}_k^H \right) \le  1, \quad k=1,2,\dots,K,
\end{align}
where $\mbf{E}_{i}\left(\mathcal{V}\op,\mathcal{B}\op,\mbf{G}_{i}\op\right)$ is the $L\times L$ MSE matrix defined in \eqref{eq:mMSEMat0_1}.

From the statistical properties of the information vectors $\mathbf{s}_i$, we can elaborate \eqref{eq:mMSEMat0_1} as
\begin{equation}
\begin{array}{l}
\mbf{E}_{i} \left(\mathcal{V}\op,\mathcal{B},\mbf{G}_{i}\op\right) = \mathbf{I}_L + \left(\mbf{G}^H_{i}\op \mathbf{H}_{i}\left(\om,\mathcal{B}\right) \mathbf{V}_i\op\right) \left(\mbf{G}^H_{i}\op \mathbf{{{H}}}_{i}\left(\om,\mathcal{B}\right) \mathbf{V}_i\op\right)^H  \\
\quad - \left(\mbf{G}^H_{i}\op \mathbf{H}_{i}\left(\om,\mathcal{B}\right) \mathbf{V}_i\op \right) - \left(\mbf{G}^H_{i}\op \mathbf{H}_{i}\left(\om,\mathcal{B}\right) \mathbf{V}_i\op \right)^H  \\
\quad + \sum\limits_{j=1 \atop j \neq i}^{N_u }\left(\mbf{G}^H_{i}\op \mathbf{H}_{i}\left(\om,\mathcal{B}\right)\mathbf{V}_j\op\right)\left(\mbf{G}^H_{i} \mathbf{H}_{i}\left(\om,\mathcal{B}\right)\mathbf{V}_j\op\right)^H + \sigma_i^2 \mbf{G}^H_{i}\op \mbf{G}_{i}\op \quad \in \mathbb{C}^{L \times L}
\end{array}
\label{eq:mMSEMat3b}
\end{equation}

The optimization problem in \eqref{wMMSE1} is constituted by an inner and an outer optimization sub-problems. To tackle both sub-problems, we employ an iterative BCD algorithm. More precisely, let us denote by $\mathcal{B}^{(q)}$, $\mbf{G}_{i}^{(q)}\op$, $\mbf{W}_{i}^{(q)}\op$ and $\mbf{V}_{i}^{(q)}\op$ the optimization variables after the $q$th iteration. As far as the inner optimization sub-problem is concerned, the set of variables $\mathcal{B}_q$ is fixed as a function of $\om$. Thus, the inner optimization sub-problem in \eqref{wMMSE1} is formally the same as the optimization problem solved during the online phase. This implies that the single-step updates $\mbf{G}_{i}^{(q+1)}\op$, $\mbf{W}_{i}^{(q+1)}\op$ and $\mbf{V}_{i}^{(q+1)}\op$ can be computed from \eqref{MMSEric}, \eqref{eq: weight computation}, and \eqref{eq:precoderShi_o}.

As far as the outer optimization sub-problem in \eqref{wMMSE1} is concerned, on the other hand, the solution is not straightforward. More precisely, the outer optimization sub-problem in \eqref{wMMSE1} encompasses the computation of the optimal set of variables $\mathcal{B}$ while all the other optimization variables are kept fixed. Next, we introduce an iterative algorithm to solve this problem. For ease of notation, we drop the index of the iterations $q$ and use the notation $\mbf{G}_{i}\left(\om\right)$, $\mbf{W}_{i}\left(\om\right)$, $\mbf{V}_{i}\left(\om\right)$.

To start with, we elaborate the expression of the MSE in \eqref{eq:mMSEMat3b}. By recalling that $\mathbf{{H}}_{i} \left(\om,\mathcal{B}\right) = \mathbf{{\bar{H}}}_{i}\left(\om\right) + \sum\nolimits_{k=1}^{K}\mathbf{\tilde{H}}_{i,k} \left(\om,\mbf{b}_{k}\right)$, we obtain
\begin{equation}
\begin{aligned}
 \mbf{E}_{i}  \left(\mathcal{V\left(\om\right)},\mathcal{B},\mbf{G}_{i}\left(\om\right)\right) & = \\
\sum\limits_{j=1}^{N_u } & \left[\mbf{G}^H_{i}\left(\om\right) \sum\limits_{k=1}^{K}\mathbf{\tilde{H}}_{i,k} \left(\om,\mbf{b}_{k}\right) \mathbf{V}_j\left(\om\right)\right] \left[\mbf{G}^H_{i}\left(\om\right) \sum\limits_{k=1}^{K}\mathbf{\tilde{H}}_{i,k} \left(\om,\mbf{b}_{k}\right) \mathbf{V}_j\left(\om\right)\right]^H  \\
& + 2 \Re \left[\sum\limits_{j=1}^{N_u }\mbf{G}^H_{i}\left(\om\right)  \left(\sum\limits_{k=1}^{K}\mathbf{\tilde{H}}_{i,k} \left(\om,\mbf{b}_{k}\right)\right) \mathbf{V}_j\left(\om\right) \mathbf{V}^H_j\left(\om\right)\mathbf{{\bar{H}}}^H_{i}\left(\om\right)\mbf{G}_{i}\left(\om\right)\right]  \\ 
& - 2 \Re \left[\mbf{G}^H_{i}\left(\om\right) \left(\sum\limits_{k=1}^{K}\mathbf{\tilde{H}}_{i,k} \left(\om,\mbf{b}_{k} \right)\right)\mathbf{V}_i\left(\om\right)\right] + \Upsilon \quad \in \mathbb{C}^{L \times L} 
\end{aligned}
\label{eq:mMSEMat4}
\end{equation}
where $2 \Re \left(\mathbf{A}\right) = \mathbf{A} + \mathbf{A}^H$, $\Upsilon$ is a constant, and $\Re (x)$ denotes the real part of $x$. 

Let us consider the WMMSE minimization with respect to the load reflection coefficient, $\mbf{b}_{m}$, of the $m$th IRS. To this end, we introduce the  shorthand notation
\begin{align}
 \mathbf{A}_{i,m}\left(\om\right) &= \mbf{G}^H_{i}\left(\om\right)  \mathbf{T}_{i,m}\left(\om\right) \quad \in \mathbb{C}^{L \times P} \nonumber \\
 \mathbf{C}_{m,j}\left(\om\right) &= \mathbf{S}_{m}\left(\om\right) \mathbf{V}_{j}\left(\om\right) \quad \in \mathbb{C}^{P \times L} \nonumber \\
  \mathbf{D}_{i,m,j}\left(\om\right) &= \mathbf{S}_{m}\left(\om\right) \mathbf{V}_{j}\left(\om\right) \mathbf{{\bar{H}}}^H_{i}\mbf{G}_{i}\op \quad \in \mathbb{C}^{P \times L} \nonumber \\
 \mathbf{F}_{i,m,j}\left(\om\right) &= \mathbf{S}_{m}\left(\om\right)\mathbf{V}_{j}\left(\om\right) \left(\mbf{G}^H_{i}\left(\om\right) \sum\limits_{k=1 \atop k \ne m}^{K} \mathbf{\tilde{H}}_{i,k} \left(\om , \mbf{b}_{k}\right)\mathbf{V}_j\left(\om\right)\right) \quad \in \mathbb{C}^{P \times L} \, .
\label{eq:definitions}
\end{align}

From \eqref{eq:rx_irs3}, \eqref{eq:mMSEMat4} and \eqref{eq:definitions}, we can derive the contribution of $\mbf{b}_{m}$ to the MSE as follows
\begin{align}
\mbf{E}_{i}^{(m)} \left(\mathcal{V}\left(\om\right),\mathcal{B},\mbf{G}_{i}\left(\om\right)\right) = & \sum\limits_{j=1}^{N_u }  \left[\mathbf{A}_{i,m}\left(\om\right) \mathbf{B}_{m} \mathbf{C}_{m,j}\left(\om\right) \right] \left[\mathbf{A}_{i,m}\left(\om\right) \mathbf{B}_{m} \mathbf{C}_{m,j}\left(\om\right) \right]^H  \nonumber \\
&+ 2 \Re \left[\sum\limits_{j=1}^{N_u }\mathbf{A}_{i,m}\left(\om\right)  \mathbf{B}_{m} \mathbf{D}_{i,m,j}\left(\om\right) \right] + 2 \Re \left[\sum\limits_{j=1}^{N_u } \mathbf{A}_{i,m}\left(\om\right) \mathbf{B}_{m}\mathbf{F}_{i,m,j}\left(\om\right)\right] \nonumber \\ 
& -2 \Re \left[\mathbf{A}_{i,m}\left(\om\right) \mathbf{B}_{m} \mathbf{C}_{m,j}\left(\om\right)\right] \quad \in \mathbb{C}^{L \times L} \, .
\label{eq:mMSEMat4.1}
\end{align}
 
For ease of writing, we introduce the compact indexing notation $\mathbf{Z}^{\left\{n\text{:}m,l\text{:}p\right\}}$, which yields the submatrix extracted from the $n$th to the $m$th rows and from the $l$th to the $p$th columns of $\mathbf{Z}$. Furthermore, let us consider the mapping $\mathbf{Q}\left(\mathbf{Q}_1,\mathbf{Q}_2\right)$ between the matrices $\mathbf{Q}_1 \in \mathbb{C}^{L \times P}$ and $\mathbf{Q}_2 \in \mathbb{C}^{P \times L}$ and the matrices $\mathbf{Q} \in \mathbb{C}^{L \times PL}$ and $\boldsymbol{\Gamma}_m \in \mathbb{C}^{PL \times L}$ defined as
\begin{equation}
\begin{array}{ll}
\mathbf{Q}^{\left\{l,(m-1)P+1\text{:}mP\right\}} (\mathbf{Q}_1,\mathbf{Q}_2)= \mathbf{Q}_1^{\left\{l,\text{:}\right\}}   \odot \left(\mathbf{Q}_2^{\left\{\text{:},m\right\}}\right)^T & \text{for} ~ l = 1,2, \ldots,L ,~ m = 1,2, \ldots,L\\
\boldsymbol{\Gamma}_m ^{\left\{(l-1)P+1\text{:}lP,l\right\}} = \mbf{b}_m & \text{for} ~ l = 1,2, \ldots,L
\end{array}
\label{eq:rx_rearrange}
\end{equation}
where $()^T$ denotes the transpose operation and $\odot$ denotes the Hadamard (element-wise) product. 

From \eqref{eq:rx_rearrange}, we obtain the identity
\begin{equation}
\begin{array}{c}
\mathbf{Q}_1 \mathbf{B}_{m} \mathbf{Q}_2 = \mathbf{Q}\left(\mathbf{Q}_1,\mathbf{Q}_2\right) \boldsymbol{\Gamma}_m
\end{array}
\label{eq:rx_irs4}
\end{equation}
from which \eqref{eq:mMSEMat4.1} can be reformulated as
\begin{align}
\mbf{E}_{i}^{(m)} & \left(\mathcal{V}\left(\om\right),\mathcal{B},\mbf{G}_{i}\left(\om\right)\right) = \nonumber \\
\sum\limits_{j=1}^{N_u } & \left[\mathbf{{Q}}\left(\mathbf{A}_{i,m}\left(\om\right),\mathbf{C}_{m,j}\left(\om\right)\right)\boldsymbol{\Gamma}_{m}\right] \left[\mathbf{{Q}}\left(\mathbf{A}_{i,m}\left(\om\right),\mathbf{C}_{m,j}\left(\om\right)\right)\boldsymbol{\Gamma}_{m}\right]^H 
\nonumber  \\ 
& +2 \Re \left[\sum\limits_{j=1}^{N_u }\mathbf{{Q}}\left(\mathbf{A}_{i,m}\left(\om\right),\mathbf{D}_{i,m,j}\left(\om\right)\right)\boldsymbol{\Gamma}_{m} \right] + 2 \Re \left[\sum\limits_{j=1}^{N_u } \mathbf{{Q}}\left(\mathbf{A}_{i,m}\left(\om\right),\mathbf{F}_{i,m,j}\left(\om\right)\right)\boldsymbol{\Gamma}_{m}\right] \nonumber  \\ 
&- 2 \Re \left[\mathbf{{Q}}\left(\mathbf{A}_{i,m}\left(\om\right),\mathbf{C}_{m,j}\left(\om\right)\right)\boldsymbol{\Gamma}_{m}\right] \quad \in \mathbb{C}^{L \times L} \, .
\label{eq:mMSEMat4.11}
\end{align}

From mathematical convenience, we introduce the notation
\begin{align}
 \mathbf{\tilde{M}}_{m}\op = \sum\limits_{i=1}^{N_u} & \alpha_i  \mathbf{W}_i\left(\om\right) \sum\limits_{j=1}^{N_u }  \mathbf{{Q}}^H\left(\mathbf{A}_{i,m}\left(\om\right),\mathbf{C}_{m,j}\left(\om\right)\right)\mathbf{{Q}}\left(\mathbf{A}_{i,m}\left(\om\right),\mathbf{C}_{m,j}\left(\om\right)\right) \quad \in \mathbb{C}^{PL \times PL} \nonumber \\
 \mathbf{\tilde{U}}^H_{m}\op  =  \sum\limits_{i=1}^{N_u} & \alpha_i  \mathbf{W}_i\left(\om\right) \sum\limits_{j=1}^{N_u } \mathbf{{Q}}\left(\mathbf{A}_{i,m}\left(\om\right),\mathbf{C}_{m,j}\left(\om\right)\right)-\mathbf{{Q}}\left(\mathbf{A}_{i,m}\left(\om\right),\mathbf{D}_{i,m,j}\left(\om\right)\right) \nonumber  \\ 
 &  + \mathbf{{Q}}\left(\mathbf{A}_{i,m}\left(\om\right),\mathbf{F}_{i,m,j}\left(\om\right)\right) \in \mathbb{C}^{PL \times L}
\label{eq:mMSEMat4.3}
\end{align}
with $\mathbf{{M}}_{m}\op  \in \mathbb{C}^{P \times P} $ and $\mathbf{{u}}_{m}\op  \in \mathbb{C}^{P \times 1} $ defined as
\begin{equation}
\begin{array}{ll}
\mathbf{{M}}_{m}\op  = \sum\limits_{l=1}^{L} \mathbf{\tilde{M}}_{m}\op ^{\left\{(l-1)P+1\text{:}lP,(l-1)P+1\text{:}lP\right\}} & \quad \text{for } l = 1,2, \ldots ,L\\
\mathbf{{u}}_{m}\op  = \sum\limits_{l=1}^{L} \mathbf{\tilde{U}}_{m}\op^{\left\{(l-1)P+1\text{:}lP,l\right\}} & \quad  \text{for } l = 1,2, \ldots ,L \, .
\end{array}
\label{eq:mMSEMat4.4}
\end{equation}

Based on these definitions, a closed-form expression for the gradient of the WMMSE in \eqref{wMMSE1} with respect to $\mathbf{b}_m$ is
\begin{equation}
\label{wMMSE2r}
\begin{aligned}
\nabla_{\mathbf{b}_m} \left[\int\limits_{\text{\boldmath{$\omega$}} \in \Omega} \sum\limits_{i = 1}^{N_{u}}\alpha_{i}  \tr\left(\mbf{W}_{i}\op\mbf{E}_{i}\left(\mathcal{V}\op,\mathcal{B},\mbf{G}_{i}\op\right)\right) f_{\Omega}\op  \right]\\
= 2\left( \, \, \int\limits_{\text{\boldmath{$\omega$}} \in \Omega}  \mathbf{{M}}_{m}\op f_{\Omega}\op\right)\mbf{{b}}_m  - 2\int\limits_{\om \in \Omega} \mathbf{{u}}_{m}\op f_{\Omega}\op
 \, .
\end{aligned}
\end{equation}

Therefore, the analog beamforming vectors of the IRSs at the $(q+1)$th iteration, $\mathbf{b}_{m}^{(q+1)}$, can be formulated in analytical form. More precisely, let $\bar{\mathbf{{M}}}_{m}^{(q)}$ and $\bar{\mathbf{{u}}}_{m}^{(q)}$ denote the statistical expectation, with respect to the random variables $\om$, of $\mathbf{{M}}_{m}^{(q)}\op$ and $\mathbf{{u}}_{m}^{(q)}\op$ evaluated at the $q$th iteration
\begin{align}
\label{wMMSE3r}
\bar{\mathbf{{M}}}_{m}^{(q)} &= \int\limits_{\text{\boldmath{$\omega$}} \in \Omega} \mathbf{{M}}_{m}^{(q)}\op f_{\Omega}\op \\
\bar{\mathbf{{u}}}_{m}^{(q)} &= \int\limits_{\text{\boldmath{$\omega$}} \in \Omega} \mathbf{{u}}_{m}^{(q)}\op f_{\Omega}\op  \nonumber \, .
\end{align}

We eventually obtain
\begin{equation}
\label{eq:precoderShi}
\mbf{b}_{m}^{(q+1)} =\left(\bar{\mbf{M}}_{m}^{(q)}+ \mu_m \right)^{-1}\bar{\mbf{u}}_{m}^{(q)}
\end{equation}
where the Lagrange multipliers $ \mu_m$ is chosen so that the \ac{GC} constraint in \eqref{wMMSE1} is fulfilled.

The proposed approach requires the analytical expression of the probability density function $f_{\Omega}\op$. If $f_{\Omega}\op$ is available in closed-form, then the integrals in \eqref{wMMSE3r} can be computed numerically. In general, however, $f_{\Omega}\op$ is seldom available in closed-form because of the large number of optimization variables involved. In this latter case, the typical approach used to overcome this issue lies in using Monte Carlo methods based on a repeated random sampling of all the optimization variables. The proposed algorithm is summarized in Algorithm \ref{Alg:DIA-PD}, where $\left \{ \om_n \right \}$, for $n = 1,\ldots,N_s$, denote the $N_s$ random samples of the Monte Carlo method.

\begin{algorithm}
\footnotesize
\caption{Proposed algorithm for (offline) sum-rate maximization}
\textbf{Initialize:}\\
Generate an initial precoding matrix ${\mathbf{V}}^{(0)}$ and IRS analog beamforming vectors ${\mathbf{b}}_k^{(1)}$, $k = 1,\ldots,K$, that fulfill the power constraint\;
Set an arbitrarily small value $\epsilon$\;
$q \leftarrow 1$, $\Delta \leftarrow 1$\;
Generate the Monte Carlo samples $\om_n$, $\forall n=1,2,\ldots, N_s$, from $f_{\Omega}\op$\\
\While{$\Delta >\epsilon $}
{
	\For {$n = 1,\ldots,N_s$}
	{
		\For {$i = 1,\ldots,N_u$}
		{
		Compute ${\mathbf{G}}_{i}^{(q)}(\om_n)$,  ${\mathbf{W}}_{i}^{(q)}(\om_n)$ and ${\mathbf{V}}_{i}^{(q)}(\om_n)$ according to \eqref{MMSEric}, \eqref{eq: weight computation} and \eqref{eq:precoderShi_o};\\ \label{algLine:gwvUpdate}
		}
		\For {$k = 1,\ldots,K$}
	        {
		$\bar{\mathbf{{M}}}_{k}^{(q)} \leftarrow 0$\\
		$\bar{\mathbf{{u}}}_{k}^{(q)} \leftarrow 0$\\

		Compute ${\mathbf{{M}}}_{k}^{(q)}(\om_n)$ and ${\mathbf{{u}}}_{k}^{(q)}(\om_n)$ according to \eqref{eq:mMSEMat4.4};\\ \label{algLine:mvUpdate}
		$\bar{\mathbf{{M}}}_{k}^{(q)} \leftarrow \bar{\mathbf{{M}}}_{k}^{(q)} + {\mathbf{{M}}}_{k}^{(q)}(\om_n)$;\\
		$\bar{\mathbf{{u}}}_{k}^{(q)} \leftarrow \bar{\mathbf{{u}}}_{k}^{(q)} + {\mathbf{{u}}}_{k}^{(q)}(\om_n);$
        		}
		}
		\For {$k = 1,\ldots,K$}
	        {
		Compute ${\mathbf{b}}_{k}^{(q+1)}$ according to \eqref{eq:precoderShi}.	
		}	

	$\Delta \leftarrow \lVert {\mathbf{b}}^{(q+1)} - {\mathbf{b}}^{(q)}\rVert$\;
	$q \leftarrow q + 1$
}
\label{Alg:DIA-PD}
\end{algorithm}

\subsection{Convergence of the proposed offline optimization algorithm}

The proposed offline optimization algorithm minimizes the average weighted MSE. In particular, all the optimization variables are updated simultaneously. From \cite{Solodov1998}, it follows that the proposed algorithm is guaranteed to converge to a stationary point of \eqref{wMMSE1}. An important property about the convergence of Algorithm \ref{Alg:DIA-PD} is stated in the following theorem.

\begin{thm}\label{conv_proof}
Any stationary point of \eqref{wMMSE1} is also a stationary point of \eqref{P:sumRateMax1}.
\end{thm}
\begin{IEEEproof}
Let $\hat{\mathcal{B}}$, $\hat{\mathcal{V}}\op$, $\hat{\mathcal{W}}\op$, and $\hat{\mathcal{G}}\op$ denote the optimization variables of Algorithm \ref{Alg:DIA-PD} at convergence. Since the WMMSE is guaranteed to converge to a stationary point of \eqref{wMMSE1}, the unconstrained variable set ${\mathcal{G}}\op$ fulfills the MMSE conditions $\mbf{E}_{i}\left(\hat{\mathcal{V}}\op,\hat{\mathcal{B}},\hat{\mbf{G}}_{i}\op\right) = \hat{\mbf{E}}_{i}\left(\hat{\mathcal{V}}\op,\hat{\mathcal{B}}\right)$ and $\hat{\mbf{W}}_{i} \op = \hat{\mbf{E}}_{i}^{-1}\left(\hat{\mathcal{V}}\op,\hat{\mathcal{B}}\right) $.
Let us define
\begin{align}
\label{Def1}
{\Theta}_1\left(\mathcal{B},\mathcal{V}\op,\mathcal{G}\op,\mathcal{W}\op\right) = &\int\limits_{\text{\boldmath{$\omega$}} \in \Omega} \sum\limits_{i = 1}^{N_{u}}\alpha_{i}  \big\{\tr\left[\mbf{W}_{i}\op\mbf{E}_{i}\left(\mathcal{V}\op,\mathcal{B},\mbf{G}_{i}\op\right)\right] \nonumber \\
& \quad \quad \quad -\log\det\left(\mbf{W}_{i}\op\right)\big\} f_{\Omega}\left(\text{\boldmath{$\omega$}}\right)  \, d\text{\boldmath{$\omega$}}  \\
 {\Theta}_2\left(\mathcal{B},\mathcal{V}\op\right) =& -\int\limits_{\text{\boldmath{$\omega$}} \in \Omega} \sum\limits_{i = 1}^{N_{u}}\alpha_{i}R_{i}\left(\mathcal{V}\op,\mathcal{B}\right) \, f_{\Omega}\left(\text{\boldmath{$\omega$}}\right) \, d\text{\boldmath{$\omega$}} \, .
\end{align}

Since the optimization problems in \eqref{wMMSE1} and \eqref{P:sumRateMax1} have the same constraints in terms of the optimization variables $\mathcal{B}$ and $\mathcal{V}\op$, Theorem \ref{conv_proof} can be proved by showing that $\Theta_1$ and $\Theta_2$ have the same gradient, computed with respect to $\mathcal{B}$ and $\mathcal{V}\op$, in correspondence of the convergence points $\hat{\mathcal{B}}$, $\hat{\mathcal{V}}\op$, $\hat{\mathcal{W}}\op$, $\hat{\mathcal{G}}\op$. In detail, with the aid of some notable results on the differentiation of complex-valued matrices, the following gradients are obtained
\begin{align}
\label{Gradients}
 \nabla_{\mbf{b}_{m}} \left\{\tr\left[ \hat{\mbf{W}}_{i} \op \hat{\mbf{E}}_{i}\left(\hat{\mathcal{V}}\op,\hat{\mathcal{B}}\right)\right]\right\} = &\hat{\mbf{W}}^T_{i} \op\nabla_{\mbf{b}_{m}} \left[\hat{\mbf{E}}_{i}\left(\hat{\mathcal{V}}\op,\hat{\mathcal{B}}\right)\right] \\
\nabla_{\mbf{b}_{m}} \left[\det\left({\hat{\mbf{E}}_{i}} \left(\hat{\mathcal{V}}\op,\hat{\mathcal{B}}\right)\right)\right]  = &  \det\left({\hat{\mbf{E}}}_{i} \left(\hat{\mathcal{V}}\op,\hat{\mathcal{B}}\right)\right) \left[\hat{\mbf{E}}^T_{i} \left(\hat{\mathcal{V}}\op,\hat{\mathcal{B}}\right)\right]^{-1} \nonumber \\
& \cdot \nabla_{\mbf{b}_{m}} \left[\left(\hat{\mbf{E}}_{i} \left(\hat{\mathcal{V}}\op,\hat{\mathcal{B}}\right)\right)\right] \nonumber   \, .
\end{align}

Since $R_{i}\left(\hat{\mathcal{V}}\op,\hat{\mathcal{B}}\right) = -\log \det\left(\hat{\mbf{E}}_{i} \left(\hat{\mathcal{V}}\op,\hat{\mathcal{B}}\right)\right)$, from \eqref{Gradients} and \eqref{Def1} we obtain
\begin{equation}
\label{Def2}
\begin{aligned}
& \nabla_{\mbf{b}_{m}} \left[{\Theta}_1\left(\hat{\mathcal{B}},\hat{\mathcal{V}}\op,\hat{\mathcal{G}}\op,\hat{\mathcal{W}}\op\right)\right] = \int\limits_{\text{\boldmath{$\omega$}} \in \Omega} \sum\limits_{i = 1}^{N_{u}}\alpha_{i} \hat{\mbf{W}}_{i}^T \nabla_{\mbf{b}_{m}} \left[\hat{\mbf{E}}_{i}\left(\hat{\mathcal{V}}\op,\hat{\mathcal{B}}\right)\right] f_{\Omega}\left(\text{\boldmath{$\omega$}}\right)d\text{\boldmath{$\omega$}}\\     
& \nabla_{\mbf{b}_{m}} \left[{\Theta}_2\left(\hat{\mathcal{B}},\hat{\mathcal{V}}\op\right)\right] = \int\limits_{\text{\boldmath{$\omega$}} \in \Omega} \sum\limits_{i = 1}^{N_{u}}\alpha_{i} \left[\hat{\mbf{E}}^T_{i} \left(\hat{\mathcal{V}}\op,\hat{\mathcal{B}}\right)\right]^{-1} \nabla_{\mbf{b}_{m}} \left[\hat{\mbf{E}}_{i} \left(\hat{\mathcal{V}}\op,\hat{\mathcal{B}}\right)\right]  f_{\Omega}\left(\text{\boldmath{$\omega$}}\right)d\text{\boldmath{$\omega$}} \, .
\end{aligned}
\end{equation}

Since $\hat{\mbf{W}}_{i} \op = \hat{\mbf{E}}_{i}^{-1}\left(\hat{\mathcal{V}}\op,\hat{\mathcal{B}}\right)$, we have $\nabla_{\mbf{b}_{m}}  \Theta_1 = \nabla_{\mbf{b}_{m}}  \Theta_2$. By using a similar procedure, it is possible to prove the equality $\nabla_{\mbf{V}_{k}} \Theta_1 = \nabla_{\mbf{V}_{k}} \Theta_2$. This concludes the proof.
\end{IEEEproof}

\section{Numerical results and discussion}
\label{sec:NumericalResults}
 \begin{figure}
 \begin{center}
\begin{tikzpicture}
\filldraw[color=black!100, fill=blue!10, thick](-8,0) rectangle (-4,4);
\filldraw[color=red, fill=red, thick](-8.02,1.5) rectangle (-7.98,2.5);
\filldraw[color=blue, fill=blue, thick](-4.02,1.75) rectangle (-3.98,2.25);
\filldraw[color=black!50, fill=black!50, thick](-6.4,1.4) rectangle (-6.2,1.6);
\filldraw[color=black!50, fill=black!50, thick](-7,2.2) rectangle (-6.8,2.4);
\filldraw[color=black!50, fill=black!50, thick](-6.3,3.2) rectangle (-6.1,3.4);
\filldraw[color=black!50, fill=black!50, thick](-6.7,0.6) rectangle (-6.5,0.8);
\filldraw[black] (-6.5,2) circle (0pt) node[anchor=west] {UEs};
\filldraw[black] (-8.8,2) circle (0pt) node[anchor=west] {IRS};
\filldraw[black] (-4.7,2) circle (0pt) node[anchor=west] {BS};

\filldraw[color=black!100, fill=blue!10, thick](-3,0) rectangle (1,4);
\filldraw[color=red, fill=red, thick](-3.02,0.86) rectangle (-2.98,0.86+0.7);
\filldraw[color=red, fill=red, thick](-3.02,0.86+0.7+0.86) rectangle (-2.98,0.86+0.7+0.86+0.7);
\filldraw[color=blue, fill=blue, thick](1.02,1.75) rectangle (-3.98+5,2.25);
\filldraw[color=black!50, fill=black!50, thick](-6.4+5,1.4) rectangle (-6.2+5,1.6);
\filldraw[color=black!50, fill=black!50, thick](-7+5,2.2) rectangle (-6.8+5,2.4);
\filldraw[color=black!50, fill=black!50, thick](-6.3+5,3.2) rectangle (-6.1+5,3.4);
\filldraw[color=black!50, fill=black!50, thick](-6.7+5,0.6) rectangle (-6.5+5,0.8);
\filldraw[black] (-6.5+5,2) circle (0pt) node[anchor=west] {UEs};
\filldraw[black] (-8.8+5,0.86+0.35) circle (0pt) node[anchor=west] {IRS};
\filldraw[black] (-8.8+5,0.86+0.7+0.86+0.35) circle (0pt) node[anchor=west] {IRS};
\filldraw[black] (-4.7+5,2) circle (0pt) node[anchor=west] {BS};

\filldraw[color=black!100, fill=blue!10, thick](-8+10,0) rectangle (-4+10,4);
\filldraw[color=red, fill=red, thick](-3.02+5,1) rectangle (-2.98+5,1+0.5);
\filldraw[color=red, fill=red, thick](-3.02+5,1+0.5+1) rectangle (-2.98+5,1+0.5+1+0.5);
\filldraw[color=red, fill=red, thick](-3+5+1,3.98) rectangle (-3+5+1.5,4.02);
\filldraw[color=red, fill=red, thick](-3+5+1,-0.02) rectangle (-3+5+1.5,0.02);
\filldraw[color=red, fill=red, thick](-3.02+5,1+0.5+1) rectangle (-2.98+5,1+0.5+1+0.5);
\filldraw[color=blue, fill=blue, thick](-4.02+10,1.75) rectangle (-3.98+10,2.25);
\filldraw[color=black!50, fill=black!50, thick](-6.4+10,1.4) rectangle (-6.2+10,1.6);
\filldraw[color=black!50, fill=black!50, thick](-7+10,2.2) rectangle (-6.8+10,2.4);
\filldraw[color=black!50, fill=black!50, thick](-6.3+10,3.2) rectangle (-6.1+10,3.4);
\filldraw[color=black!50, fill=black!50, thick](-6.7+10,0.6) rectangle (-6.5+10,0.8);
\filldraw[black] (-6.5+10,2) circle (0pt) node[anchor=west] {UEs};
\filldraw[black] (-8.8+10,0.86+0.35) circle (0pt) node[anchor=west] {IRS};
\filldraw[black] (-3+5+0.9,4.3) circle (0pt) node[anchor=west] {IRS};
\filldraw[black] (-8.8+10,0.86+0.7+0.86+0.35) circle (0pt) node[anchor=west] {IRS};
\filldraw[black] (-3+5+0.9,-0.3) circle (0pt) node[anchor=west] {IRS};
\filldraw[black] (-4.7+10,2) circle (0pt) node[anchor=west] {BS};
\end{tikzpicture}
\caption{\footnotesize{Layout of the considered indoor scenario (aerial view) for $N_u$ = 4 \acp{UE} and $N_{IRS} = 1,2, 3,4$.}}
\label{fig_scenario}
\end{center}
\end{figure}
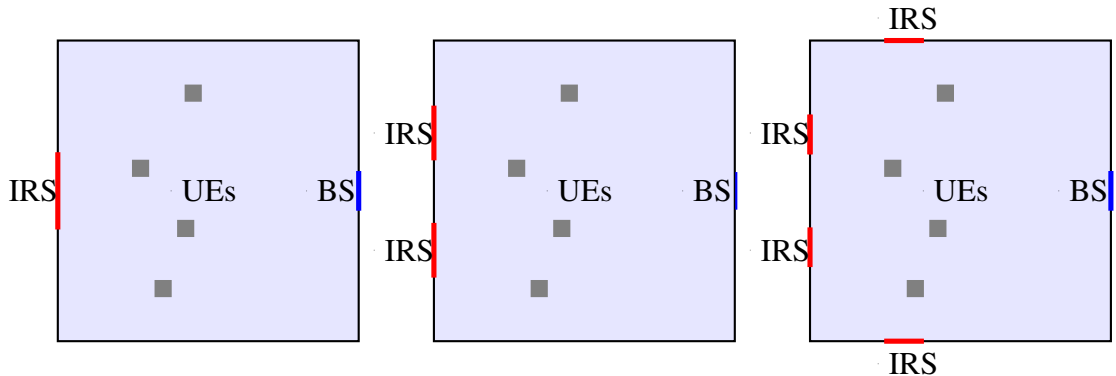
In this section, we illustrate some numerical results in order to analyze the performance of the proposed optimization algorithm, and, in particular, to assess the impact of not using instantaneous CSI for optimizing the IRSs.

\subsection{Simulation scenario}
We consider an indoor environment with dimensions $30$ m  $\times$ $30$ m along the $x$-axis and $y$-axis, respectively, as depicted in Fig. \ref{fig_scenario}. The \acp{UE} are equipped with $L = 4$ vertically polarized antennas and are deployed randomly on the left-half of the room, i.e., in an area of $15$ m $\times$ $30$ m, at a height of $1$ m. The \ac{BS} is equipped with $M = 16$ vertically polarized antennas and is placed on the opposite side of the room at the position $(15,30)$ m and at a height of $2$ m. The $16$ antennas of the \ac{BS} are arranged on the $y-z$ plane with 8 antennas along the $y$-axis and 2 antennas along the $z$-axis, while the 4 antennas of the \acp{UE} are arranged horizontally along the $y$-axis. The antenna-element gains at the \ac{BS} and at the \ac{UE} are set to 3 dB. The variance $\sigma^{2}_{i}$ of the additive zero-mean Gaussian noise and the maximum power budget $P_{i}$ (assumed the same for all UEs), are set equal to $-97\,$dBm and $0\,$dBm, respectively.  

\subsection{IRS deployment}
The \acp{IRS} are placed on the walls around the \acp{UE}. The size of each IRS is $0.428 / \sqrt{N_{IRS}}$ m $\times$ $0.214 / \sqrt{N_{IRS}}$ m. 
This ensures that, regardless of the number of IRSs, the total area covered by the IRSs is $0.428 $ m $\times$ $0.214$ m, for a fair comparison. The inter-distance between the scattering elements of the \ac{IRS} is $\lambda/2$, where $\lambda = 0.011$ m corresponds to a carrier frequency of $28$ GHz. Accordingly, the total number of scattering elements that is available in the considered environment is $P_{IRS}=3200$. To ensure an affordable computational complexity for Algorithm \ref{Alg:DIA-PD}, each \ac{IRS} is partitioned into $K_t = 64$ tiles each comprising  $P = 50$ scattering elements. As a matter of fact, the complexity of the offline optimization algorithm is mainly determined by the computation of $\mbf{b}_{m}^{(q+1)}$ in \eqref{eq:precoderShi}, which requires the inversion of a $P \times P$ matrix whose complexity is $O(P^3)$. As a function of $K_t$, on the other hand, the complexity of Algorithm \ref{Alg:DIA-PD} increases only linearly. As for the power constraints of the IRSs, both \ac{GC} and \ac{LC}$_{N_b}$ case studies are considered. 

\subsection{Channel model}
As for the wireless channel, we assume an \ac{NLOS} channel model for the direct path, i.e., $X_d =0$ in \eqref{eq:chmodel1}. This represents the most interesting network scenario for using the IRSs. If a strong \ac{LOS} path is available, in fact, the impact and contribution of the IRSs are usually less significant if the transmission distances are large and the IRSs have a small size \cite{Perovic}. As for the \ac{NLOS} contribution to the channel, we consider $N_{c} = 5$ and $N_{p} = 10$. More specifically, the centers of the clusters are positioned on an ellipse as discussed in Section \ref{direct_link} with an eccentricity $\epsilon = 0.5$, and the $N_{p}$ scatterers associated to each cluster are randomly distributed around the clusters' centers such that the angle spread of the signals received by the \acp{UE} is 15$^\circ$. These values are in agreement with typical \ac{NLOS} multipath models for \ac{mmWave} communications, e.g., see \cite{Buzzi2018}. 

\subsection{IRS optimization}
In order to assess the effectiveness of the proposed joint online-offline optimization algorithm, we compare the sum-rate of two network setups: (i) the IRSs are optimized according to Algorithm \ref{Alg:DIA-PD} (denoted by OPT) and (ii) the IRSs act as non-controllable scatterers (denoted by NON-OPT). In this latter case, the unit cells of the IRSs operate as diffusive radiating elements whose reflection coefficient has unit amplitude and whose phases are independently and uniformly distributed in $[0,2\pi]$. Also, the weights $\boldsymbol{\alpha}$ in \eqref{P:sumRateMax_pre} are set equal to 1.

As for the statistics of the locations of the \acp{UE}, we consider three scenarios that are denoted by $UD$, $UD$-1m, and $UD$-0m. In the $UD$ scenario, the \acp{UE} are deployed uniformly at random within a given service area. This corresponds to the scenario with the minimal a priory information about the locations of the UEs, and, therefore, the analog beamforming vectors of the IRSs are optimized by averaging over a probability density function $f_{\Omega}\op$ that accounts for the largest uncertainty. In the $UD$-1m scenario, the locations of the \acp{UE} are assumed to be known with an uncertainty of $0.5$ m around the actual location of each UE. In other words, the UEs are distributed within a disk of diameter $1$ m that is centered at their actual locations. In the $UD$-0m scenario, the locations of the \acp{UE} are assumed to be perfectly known. This case study corresponds to the scenario in which the maximal a priori CSI is necessary and the IRSs need to be reconfigured more often based on the UEs' locations. It is worth recalling that the IRS-assisted links are determined only by the positions of the UEs. Therefore, the $UD$ scenario corresponds to the full CSI setup for optimizing the analog beamforming vectors of the IRSs.

The simulation results are obtained by averaging the sum-rate over 10,000 network realizations and by setting $N_s = 1,000$ in Algorithm \ref{Alg:DIA-PD}. 

\begin{figure}[!h]
	\centering
	\includegraphics[width=0.5\linewidth]{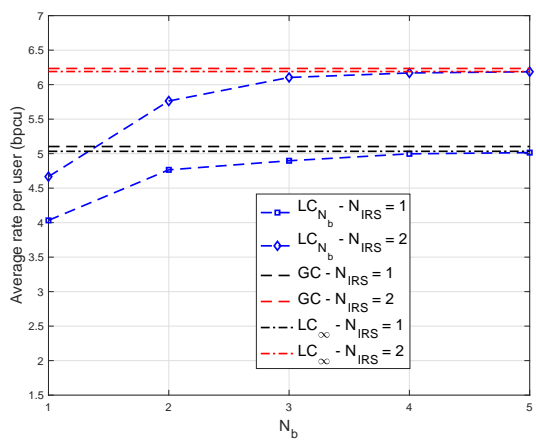}
	\caption{\footnotesize{Performance of the OPT scheme for the \ac{LC}, \ac{GC}, and \ac{LC}$_\infty$ constraints as a function of $N_b$ ($N_{IRS} = 1,2$ and $N_u = 1$).}}
	\label{F1}	
	\vskip 0.5cm
\end{figure}

\subsection{Local vs. global power constraint and phase quantization noise}
In Fig. \ref{F1}, we investigate the impact of the phase quantization noise on the sum-rate, by assuming a single UE scenario ($N_u = 1$), the OPT setup, and the IO channel model. The constraints \ac{LC}$_{N_b}$, \ac{GC}, and \ac{LC}$_\infty$ are considered, where the latter constraint is referred to the case study with no phase quantization noise. We observe that the \ac{LC} constraint provides, in the considered case study, similar performance as the \ac{GC} constraint, even if the number of quantization bits $N_b$ is low. For example, the performance difference is less than 2$\%$ if $N_b = 3$. Similar results are obtained for all considered network scenarios. For simplicity, therefore, only the curves that correspond to the \ac{GC} constraint are reported in the following figures.

\subsection{Sum-rate optimization}
In Fig. \ref{F3}-(left), we compare the OPT and NON-OPT configurations of the IRSs, by assuming the $UD$-1m scenario and the IO channel model. The curves show the sum-rate as a function of the number of users $N_u$ for $N_{IRS} = 1$, $N_{IRS} = 2$, and $N_{IRS} = 4$. We observe that the OPT configuration outperforms the NON-OPT configuration, which substantiates the effectiveness of deploying IRSs even if the locations of the UEs are not perfectly known.
It is worth noting that the average rate per user decreases with the increase of the number of \acp{UE}. This can be attributed to a stronger interference in the presence of multiple UEs. 
As far as the impact of the number of \acp{IRS} is concerned, we observe that, in general, the average rate per user increases with $N_{IRS}$ even if the total surface area covered by the IRSs is kept unchanged, i.e., deploying a larger number but smaller in size IRSs yields better performance. This is attributed to the fact that an higher number of IRSs located in different positions allow to increase the multipath, and, hence, the rank of the channel matrices as shown in subsequent results.

Slightly different results are, on the other hand, obtained in the SM channel model, as illustrated in Fig. \ref{F3}-(right). In this case, in fact, the contribution of the NLOS multipath undergoes a smaller path-loss, and, therefore, fewer IRSs may be needed for achieving good performance. 

In Fig. \ref{F5}, we show the average rate per user that is achievable in the $UD$-0m and $UD$ scenarios by assuming the IO channel model. By comparing Fig. \ref{F3} with Fig. \ref{F5}, we observe, as expected, that the $UD$-0m setup outperforms the $UD$-1m setup, but at the cost of reconfiguring the IRSs more frequently. Likewise, the $UD$ scenario offers the worse performance but with the main advantage of reducing the rate at which the IRSs need to be reconfigured. Nevertheless, it is interesting to note that the OPT case still outperforms the NON-OPT case even if the locations of the UEs are known with a quite large uncertainty.

\begin{figure}
     \centering
     \begin{subfigure}[b]{0.45\textwidth}
         \centering
         	\includegraphics[width=\linewidth]{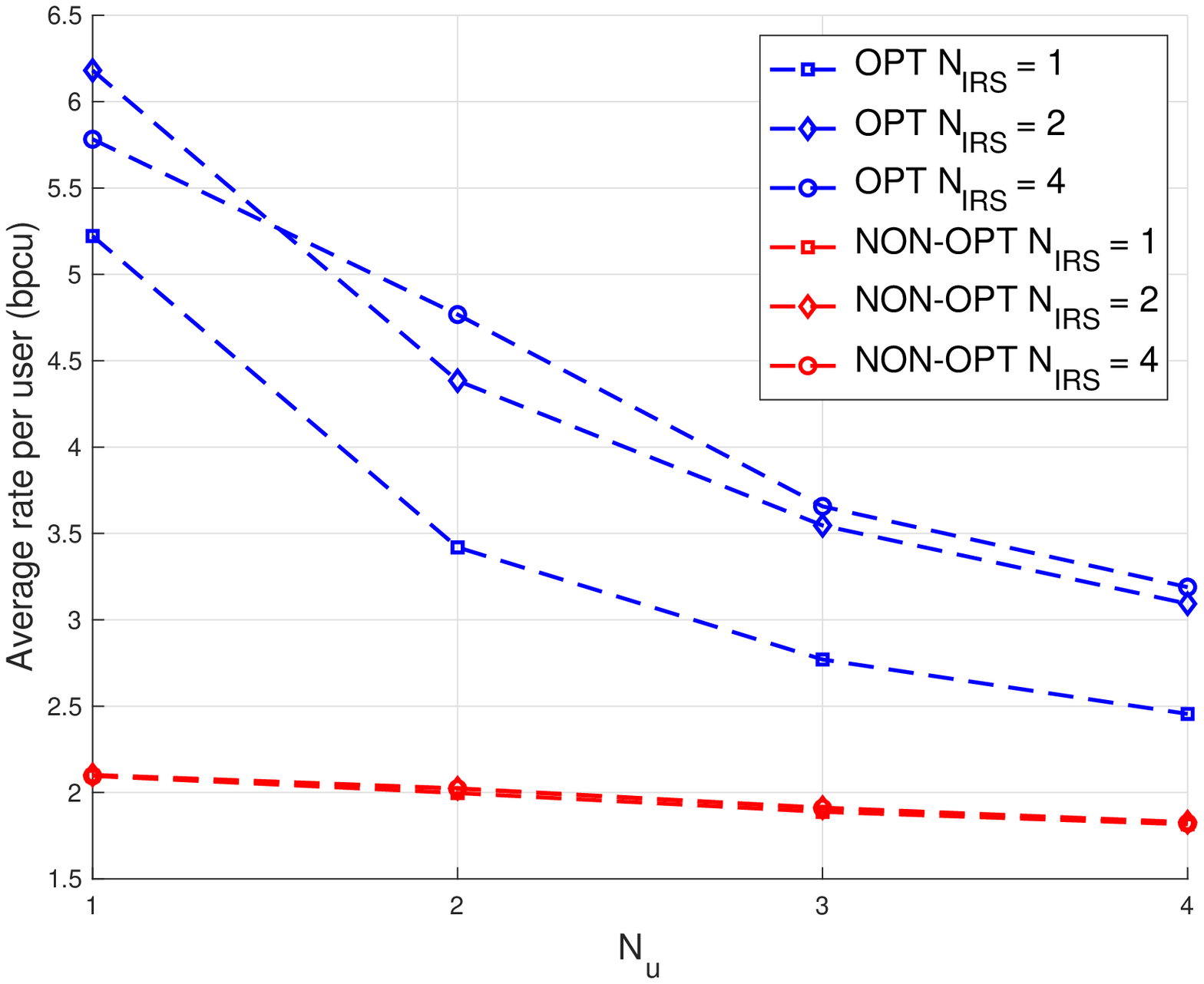}
     \end{subfigure}
     \hfill
     \begin{subfigure}[b]{0.45\textwidth}
         \centering
        \includegraphics[width=\linewidth]{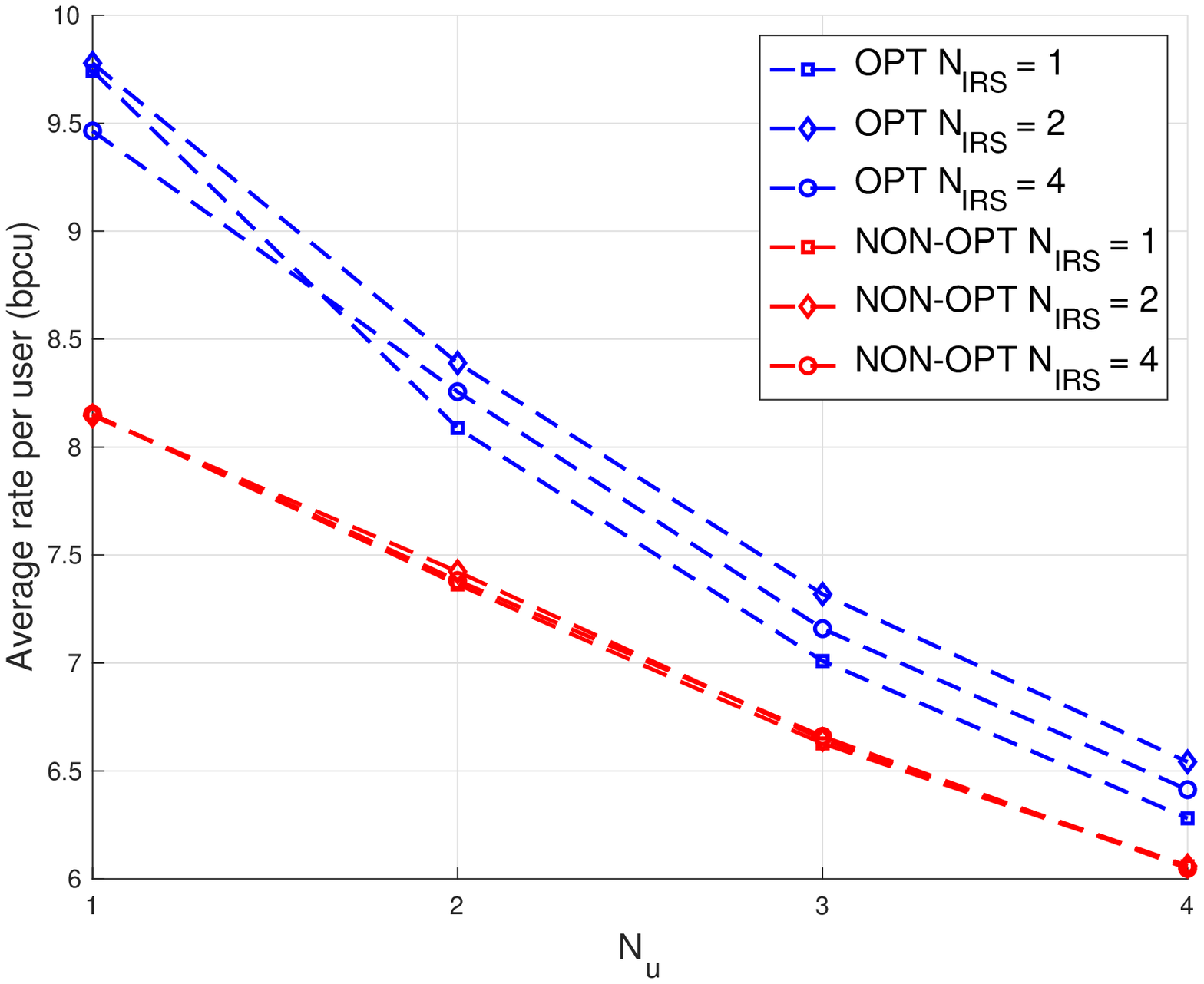}
     \end{subfigure}
     \caption{\footnotesize{Comparison between the OPT and NON-OPT schemes. (left) $UD$-1m setup and IO channel model; (right) $UD$-1m setup and SM channel model.}} 
        \label{F3}
        \vskip 1cm
\end{figure}

\begin{figure}
     \centering
     \begin{subfigure}[b]{0.45\textwidth}
         \centering
         	\includegraphics[width=\linewidth]{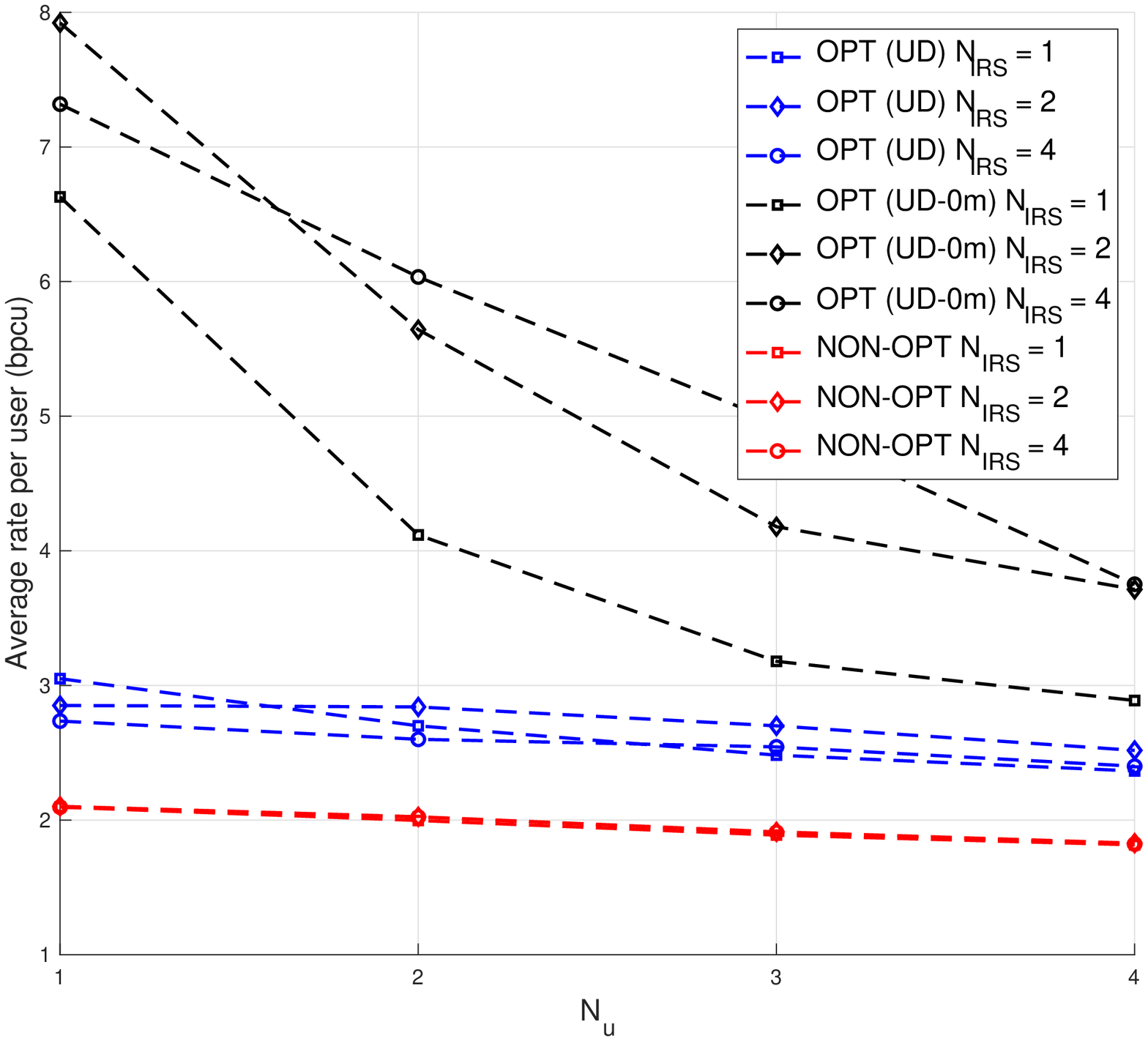}
     \end{subfigure}
     \hfill
     \begin{subfigure}[b]{0.45\textwidth}
         \centering
        \includegraphics[width=\linewidth]{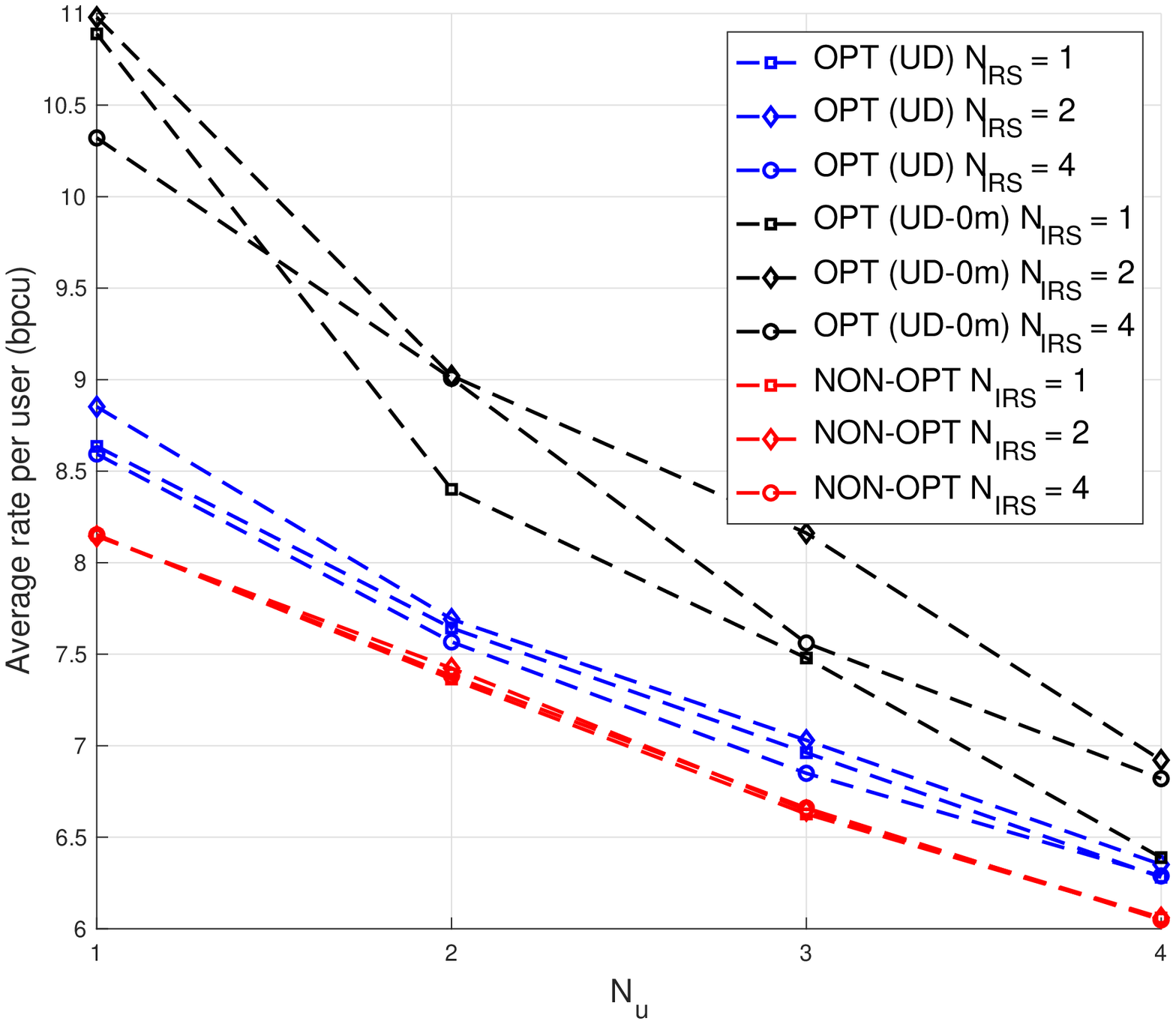}
     \end{subfigure}
     \caption{\footnotesize{Comparison between the OPT and NON-OPT schemes. (left) $UD$-1m vs. $UD$ setups and IO channel model; (right) $UD$-0m vs. $UD$ setups and IO channel model.}}
        \label{F5}
        \vskip 1.5cm
\end{figure}

\subsection{Rank of the IRS-assisted channel}
In Fig. \ref{F8}, we analyze the effective rank of the IRS-assisted channel, by assuming $N_u = 1$ and the OPT configuration of the IRSs in the three scenarios $UD$, $UD$-1m, $UD$-0m. Given the channel matrix in \eqref{eq:rx_total}, we compute $R = \sum |\lambda_i| / \max |\lambda_i|$ as a function of the number of \acp{IRS}, where $\lambda_i$ are the singular values of the channel matrix in \eqref{eq:rx_total}. The parameter $R$ provides one with information on how well conditioned the IRS-assisted channel is. In other words, the larger $R$ is, the larger the capacity and the number of information streams that can be transmitted simultaneously are. In particular, $R=1$ indicates that a singular value is predominant, and, thus, the rank of the channel is, in practice, one. If, on the other hand, all singular values are almost the same, the channel capacity is maximized. In the considered system setup, $R \le 4$, since four antennas are available at the UEs. The results in Fig. \ref{F8} show that increasing the number of IRSs increases the effective rank of the channel and, thus, the system sum-rate. The effective rank is larger in the SM channel model because of the more favorable propagation conditions.

\begin{figure}
     \centering
     \begin{subfigure}[b]{0.45\textwidth}
         \centering
         	\includegraphics[width=\linewidth]{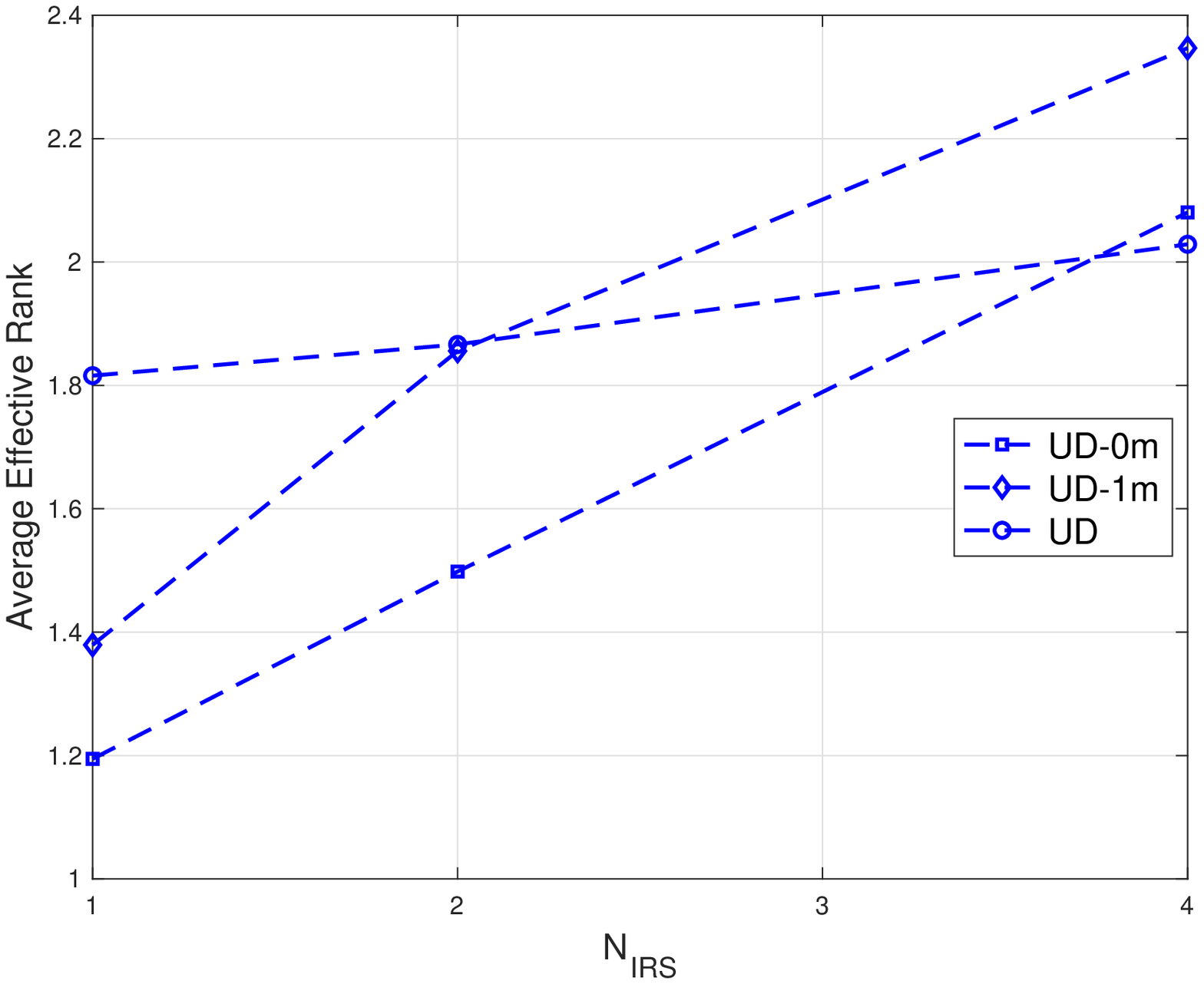}
     \end{subfigure}
     \hfill
     \begin{subfigure}[b]{0.45\textwidth}
         \centering
        \includegraphics[width=\linewidth]{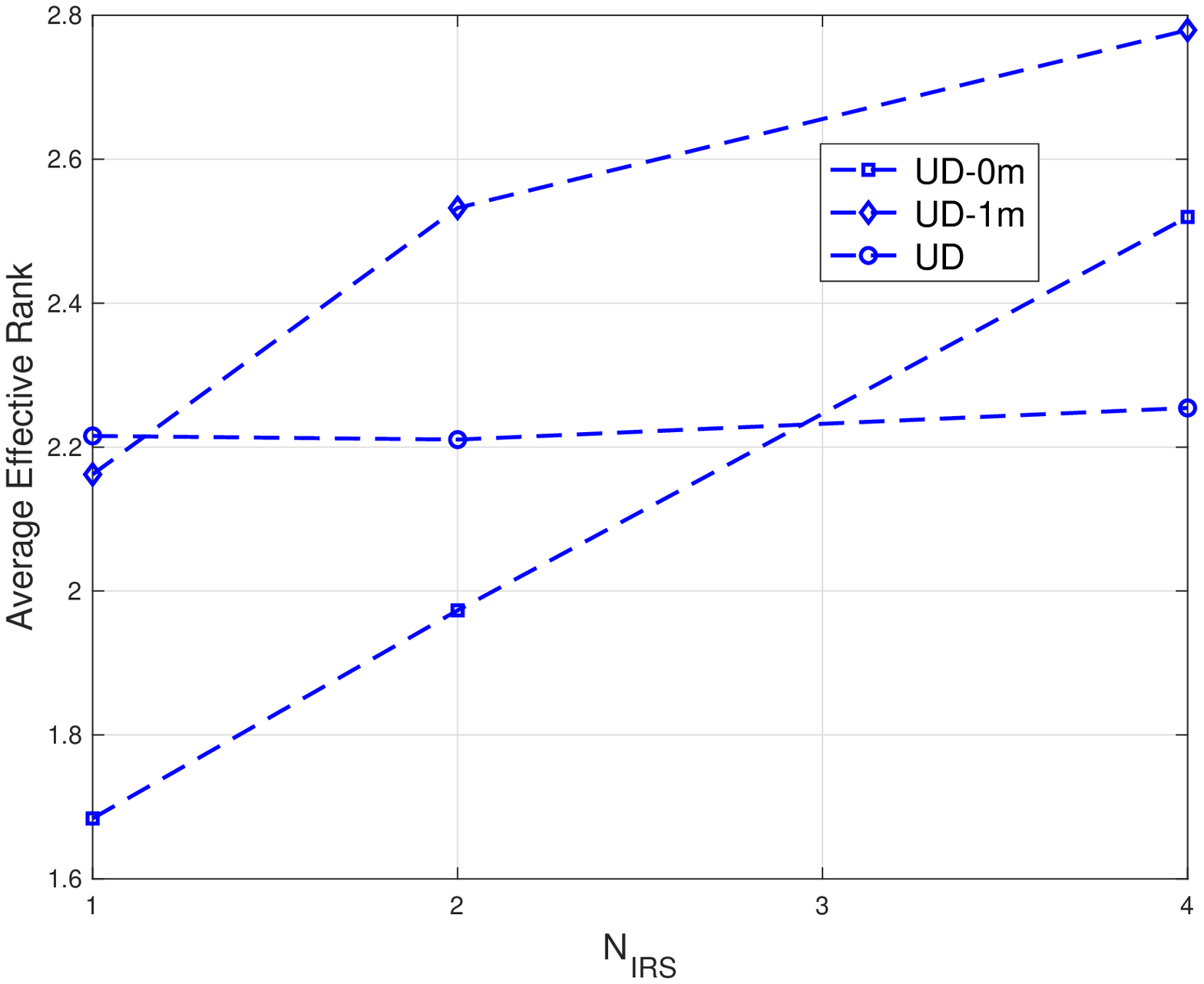}
     \end{subfigure}
     \caption{\footnotesize{Average effective rank $E(R)$ of the IRS-assisted channel as a function of $N_{IRS}$ for the $UD$, $UD$-1m, $UD$-0m setups and $N_u = 1$. (left) IO channel model; (right) SM channel model.}}
        \label{F8}
        \vskip 0.5cm
\end{figure}

\subsection{Analysis of the equivalent array factors of the IRSs}
In order to get engineering insights on how the IRSs shape the wireless channel for increasing the sum-rate, Figs. \ref{fig:AreaCircolare1m_NUE2_NIRS4} and \ref{fig:AreaUniforme_NUE4_NIRS2} report the array factor of the BS and the equivalent array factor of the IRSs for some given locations of the UEs. More precisely, the equivalent array factor of an IRS, corresponding to some given locations of the UEs, is obtained by considering an equivalent antenna array virtually located at the IRS and whose array factor is obtained from the digital beamforming vector of the BS, the channel response of the \ac{BS}-\ac{IRS} link, and the analog beamforming vector of the IRS \cite{DarMas:20}. It is worth noting that the equivalent array factor of each \ac{IRS} may change even if the IRSs are computed during the offline phase and are not updated during the online phase. This is because the array factor of the \ac{BS}, the wireless channels, and the locations of the UEs are different during the online phase.

In Fig. \ref{fig:AreaCircolare1m_NUE2_NIRS4}, we illustrate the obtained equivalent array factors by assuming $N_{IRS}=4$ and $N_u=2$ (left figure) and  $N_{IRS}=2$ and $N_u=4$ (right figure) in the $UD$-1m scenario. The equivalent array factors correspond to a single stream emitted by the BS. The equivalent array factors of the IRSs that correspond to each UE in the system are depicted in a different color. 
When the number of \acp{IRS} is larger than the number of UEs (left figure), the proposed optimization algorithm leads to a configuration in which each UE receives the signals reflected by the two closest \acp{IRS}. In fact, we note  that the equivalent array factor of the \ac{BS} associated to each UE points towards two \ac{IRS}s, which, in turn, reflect the signal towards a single UE. When the number of UEs is larger than the number of IRSs (right figure), on the contrary, the proposed optimization algorithm configures the IRSs in such a way that they reflect the signals mainly towards their respective nearest UEs (UE-1 and UE-2 in the figure), whereas the farthest UEs (UE-3 and UE-4 in the figure) are served directly by the \ac{BS} through the uncontrolled multipath from the environment.  

\begin{figure}
     \centering
     \begin{subfigure}[b]{0.48\textwidth}
         \centering
         	\includegraphics[width=\linewidth]{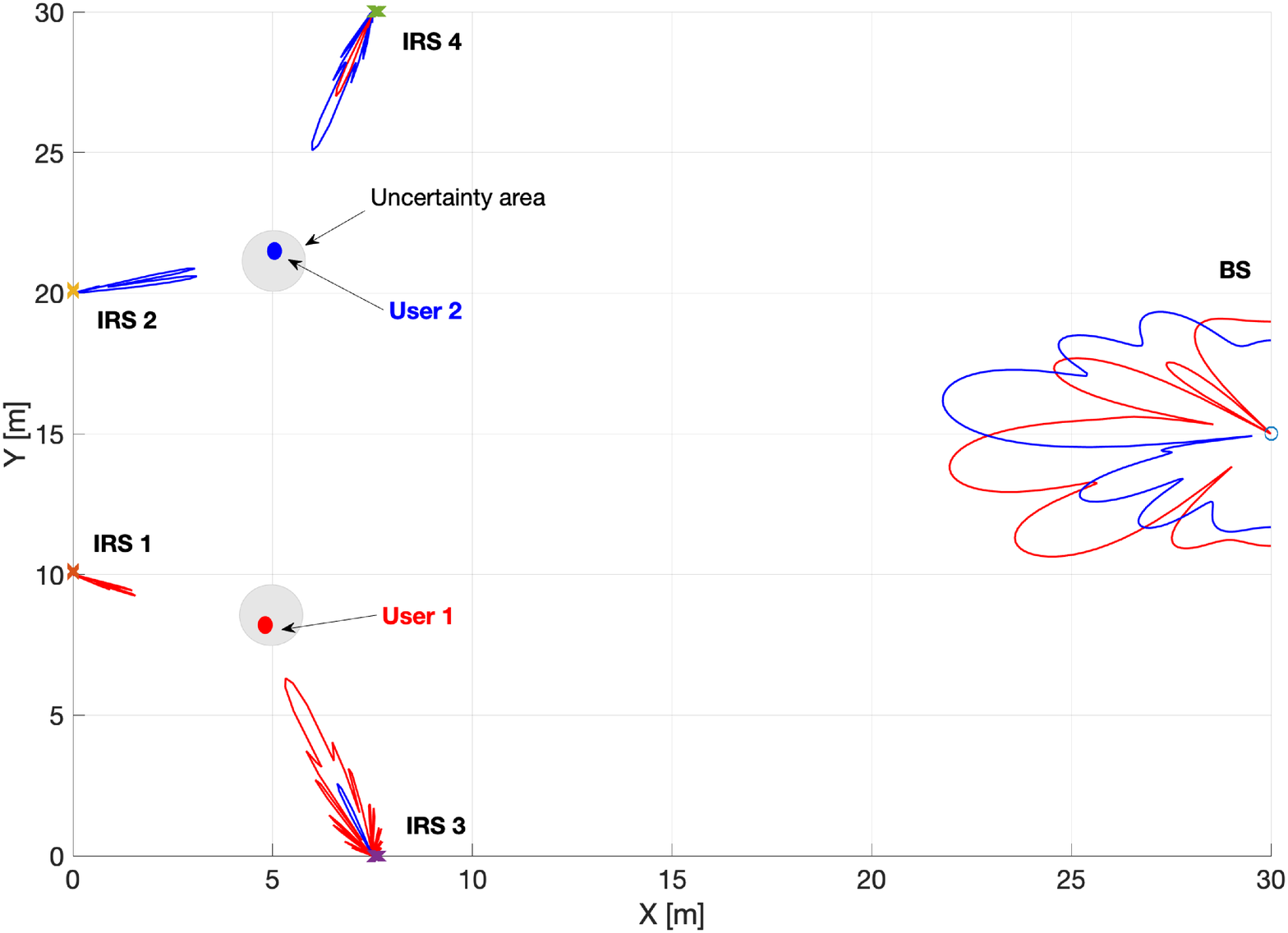}
     \end{subfigure}
     \begin{subfigure}[b]{0.48\textwidth}
         \centering
        \includegraphics[width=\linewidth]{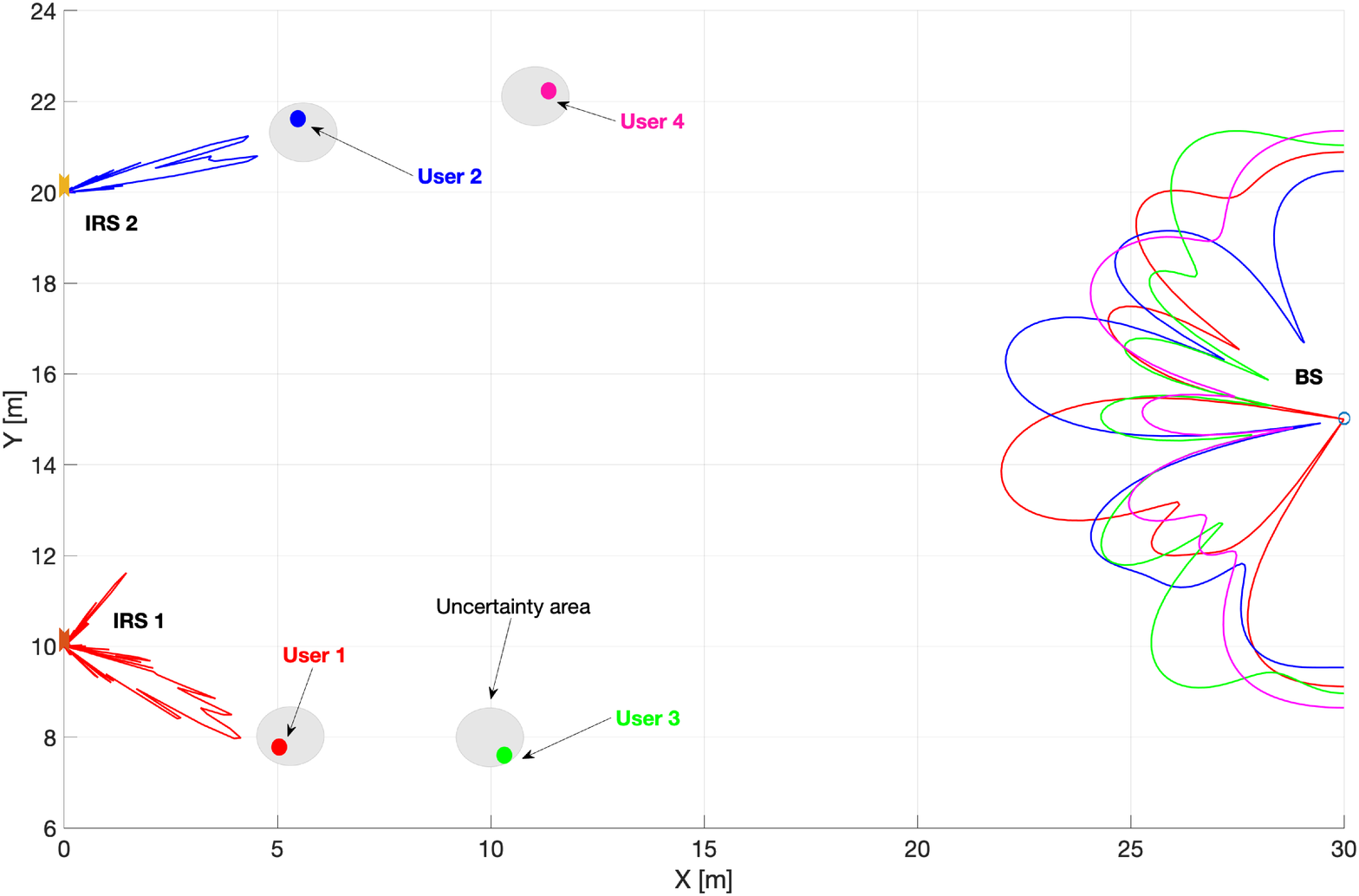}
     \end{subfigure}
     \caption{\footnotesize{BS and IRS equivalent array factors. (left) $N_{IRS}=4$, $N_u=2$, and $UD$-1m; (right) $N_{IRS}=2$, $N_u=4$, and $UD$-1m.}}
     \label{fig:AreaCircolare1m_NUE2_NIRS4}
        \vskip 0.5cm
\end{figure}

\begin{figure}[t]
	\vskip 1cm
	\centering
	\includegraphics[width=0.6\linewidth]{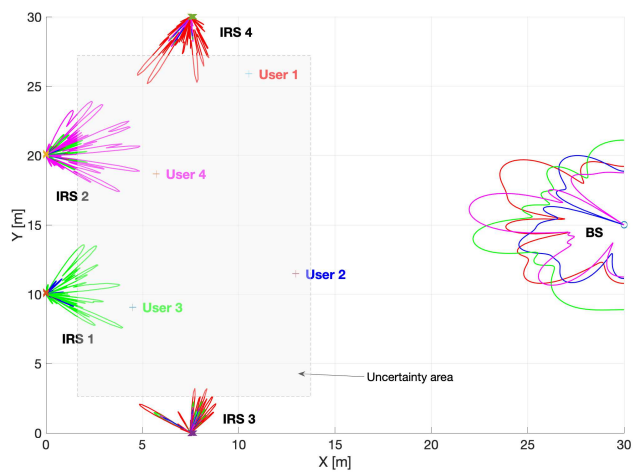}
	\caption{\footnotesize{BS and IRS equivalent array factors for $N_{IRS}=4$ and $N_u=4$. The location uncertainty of the UEs is the grey region.}}
	\label{fig:AreaUniforme_NUE4_NIRS2}
	\vskip 0.5cm
\end{figure}

Finally, Fig. \ref{fig:AreaUniforme_NUE4_NIRS2} shows the equivalent array factors in the $UD$ scenario. In this case, in particular, the IRSs are optimized with the only a priori information that the UEs can be located anywhere in the grey shadowed area (uncertainty region). The figure reports the results that correspond to the system setup $N_{IRS}=4$ and $N_u=4$ for a generic realization of the positions of the \acp{UE}.
In this case, we observe that the \acp{IRS} are optimized in order to diffuse the reflected signals towards nearby UEs while, at the same time, reducing the interference towards more distant UEs. Let us consider, for instance, UE-3 (in green) and UE-4 (in magenta). We observe that they are primarily served by IRS-1 and IRS-2, respectively. As far as UE-3 and UE-4 are concerned, in addition, we observe that IRS-3 and IRS-4 are configured in such a way they create little interference towards them, i.e., the equivalent array factors of IRS-3 and IRS-4 have notches towards the directions of UE-3 and UE-4. It is also interesting to note that UE-2 is, on the other hand, mainly served by the nearby BS through the uncontrolled multipath from the environment.

\section{Conclusion}
\label{sec:Conclusion}

In  this  paper,  we have introduced an optimization algorithm in order to maximize the network sum-rate of a MU-MIMO system in the presence of multiple \acp{IRS}. The proposed algorithm consists of an offline phase and an online phase. During the offline phase, the \acp{IRS} are configured based on an optimization procedure that relies exclusively on the statistical characterization  of  the  environment,  such  as  the  distribution   of  the  users' locations  and the  channel  statistics. In this phase, in particular, instantaneous CSI is not necessary.  During the online phase, on the other hand, only the beamforming vectors at the BS and at the users are optimized, whereas the configuration of the IRSs is kept unchanged. The main advantage of the proposed approach lies in not requiring the estimation of the instantaneous CSI for optimizing the \acp{IRS}, thus significantly  reducing the overhead associated with the optimization of IRS-assisted wireless systems.  
The obtained numerical results confirm the validity of the proposed approach. Notably, IRS-assisted wireless systems that are optimized based solely on long-term CSI still provide large performance gains as compared to wireless systems in the absence of IRSs.

\bibliographystyle{IEEEtran}
\bibliography{IEEEabrv,main2}

\end{document}